\documentclass[10pt,journal,letterpaper,compsoc]{IEEEtran}
\usepackage{subfig}
\usepackage{amsmath}
\usepackage{amsthm,algorithmic,graphicx}
\usepackage[boxed,commentsnumbered,vlined,ruled,linesnumbered]{algorithm2e}
\usepackage{multirow}
\usepackage{xcolor}
\usepackage{verbatim}
\usepackage{float}
\usepackage{amssymb}
\usepackage{hyperref}
\usepackage[normalem]{ulem}
\usepackage{array,booktabs}
\usepackage{bibunits}
\usepackage{soul}
\usepackage{url}
\usepackage{tikzpagenodes}
\usepackage{xcolor}
\usepackage{tabularx}
\newcolumntype{L}{>{\raggedright\arraybackslash}X}
\usepackage{blindtext}
\usepackage{enumitem}
\usepackage{adjustbox}
\usepackage{diagbox}
\usepackage{graphicx}
\newsavebox{\bigimage}

\usepackage{array,multirow,graphicx}
\newcolumntype{R}[2]{%
    >{\adjustbox{angle=#1,lap=\width-(#2)}\bgroup}%
    l%
    <{\egroup}%
}

\graphicspath{ {./figures/} }

\begin{document}

\title{Fleet management for ride-pooling with meeting points at scale: a case study in the five boroughs of New York City}

\author{Motahare Mounesan, Vindula Jayawardana, Yaocheng Wu, Samitha Samaranayake, Huy T. Vo
\IEEEcompsocitemizethanks{
\IEEEcompsocthanksitem Motahare Mounesan and Yaocheng Wu are with the Graduate Center, City University of New York, New York, USA.

\IEEEcompsocthanksitem Vindula Jayawardana is with the Laboratory for Information and Decision Systems, Massachusetts Institute of Technology, Cambridge, USA.

\IEEEcompsocthanksitem Samitha Samaranayake is with the School of Civil and Environmental Engineering, Cornell University, Ithaca, USA.

\IEEEcompsocthanksitem Huy T. Vo is with the City College, City University of New York, and the Center for Urban Science and Progress, New York University, New York, USA.
}
}


\IEEEcompsoctitleabstractindextext{
\begin{abstract}
Introducing meeting points to ride-pooling (RP) services has been shown to increase the satisfaction level of both riders and service providers. 
Passengers may choose to walk to a meeting point for a cost reduction. 
Drivers may also get matched with more riders without making additional stops. There are 
economic benefits of using ride-pooling with meeting points (RPMP) compared to the traditional RP services. Many 
RPMP models
have been proposed to better understand their benefits. However, most prior works 
study RPMP either with a restricted set of parameters
 or at a small scale due to the expensive computation involved. In this paper, we propose STaRS+, a scalable RPMP framework that is based on a comprehensive integer linear programming model. The high scalability of STaRS+ is achieved by utilizing a heuristic optimization strategy along with a novel shortest-path caching scheme. We applied our model to the NYC metro area to evaluate the scalability of the framework and demonstrate
 the importance of city-scale simulations.
 Our results show that city-scale simulations can reveal valuable insights 
 for city planners 
 that are not always visible at 
 smaller scales.
  To the best of our knowledge, STaRS+ is the first study on the RPMP that can solve large-scale instances on the order of the entire NYC metro area.
\end{abstract}
\begin{IEEEkeywords}
ride-sharing, scalability, shortest-path caching, simulation.  
\end{IEEEkeywords} 
}

\maketitle
\IEEEdisplaynotcompsoctitleabstractindextext
\IEEEpeerreviewmaketitle


\section{Introduction}

\IEEEPARstart{R}{ide-hailing} services have become extremely popular in urban areas by
providing a convenient and reliable mode of travel for many. 
They are extremely easy to use and are available at a price point that is competitive against traditional taxi services. While ride-hailing offers a number of benefits, these services also impose a significant burden on the transportation network due to their inefficiencies. More precisely, the asymmetry between supply and demand results in ride-hailing vehicles require the fleet to be constantly re-positioned (or \textit{rebalanced)}, to ensure that the vehicle locations are aligned with user
demand. Therefore, in contrast to these trips being taken via a personal vehicle, rebalancing has the unfortunate side-effect of increasing the total vehicle mileage
driven in the system to satisfy the same demand. This raises concerns
about increasing congestion on city streets, as has been expressed
recently by multiple city planners including NYC~\cite{CongestionPricing1,CongestionPricing2}. Given
that ride-hailing vehicles can carry multiple passengers
simultaneously, one strategy for mitigating this effect is to promote
ride-pooling (RP) into the service, encouraging passengers to share rides
with other commuters. Most major ride-hailing companies already
provide a pooling service (e.g. UberPool, Lyft Line, and Via), where
passengers pay a reduced fee to compensate for the inconveniences
(additional travel time, sharing a vehicle etc.) associated with
pooling. More recently, there have also been more transit like
ride-pooling services that require users to walk to pickup points (and
from dropoff points) instead of providing a door-to-door service
(e.g. Uber Express Pool, Lyft Shared Saver). This can further increase
the efficiency of the ride-hailing system and reduce its negative
externalities.

There are many factors that contribute to the efficiency of
ride-hailing services arising from various, but often conflicting,
interests of stakeholders including city planners, service providers,
drivers, as well as customers. In fact, a large body of recent work has
been invested in door-to-door ride-pooling simulations to better understand
these factors. However, only a limited number of these studies tackle the related problem of door-to-door ride-pooling with meeting points (RPMP), where passengers can be asked to walk to a meeting point. This is mostly because
of the expensive computational overhead involved. In addition, most previous
works, if not all, only focus on small-scale scenarios, such as using
only part of a city as their testbeds. For example, despite a large
number of rail-hailing studies in New York City (NYC), to the best of
our knowledge, none had been conducted for the entire NYC metro
area. Instead, many chose to study only Manhattan (one of the five
boroughs in NYC) to ease the computational challenge. However, these
results might not generalize
to the entire city due to the unique
characteristics of Manhattan. To address this issue, we present a
scalable framework that allows exploration of ride-sharing scenarios
at full-city scale. In particular, our contributions are as follows:

\begin{itemize}[leftmargin=*]
\item A comprehensive RPMP model with an implementation using integer linear programming (ILP).

\item A scalable framework to simulate the proposed RPMP
  model using an iterative heuristic optimization and a novel shortest path caching scheme.

\item A first RP study that is capable of operating on the entire NYC metro area, with detailed comparisons of our heuristic approach against the baseline ILP model.
\end{itemize}


\section{Related Works}
The problem of designing dynamic matching algorithms for RP has an extensive literature springing from work on supply-chain and vehicle routing problems~\cite{gendreau1998dynamic} and is closely related to the problem of capacitated vehicle routing with time windows~\cite{Des98} and dynamic pickup and delivery~\cite{BERBEGLIA20108}. 
Due to the dynamic nature of the problem, there is often a tradeoff between the optimality and tractability of the proposed solutions. Though Integer Linear Programming (ILP) naturally fits the problem formulations, heuristic methods are required for scalability. 

For large-scale simulation, Ma et al. \cite{ma2013t} proposed a heuristic-based (dynamic) taxi dispatching strategy for large urban fleets alongside a pricing scheme. The set of candidate taxis can serve the query first retrieved by a dual-side searching algorithm utilizing a novel spatio-temporal vehicle indexing scheme. Next, a scheduler algorithm finds the optimal solution.
Santi et al.~\cite{santi2014quantifying} formulated the static spatio-temporal sharing problem using a graph-theoretic framework,  introducing the notion of a shareability network, and quantitatively assessed the impact of large-scale sharing for capacity two vehicles.
Ota et al.~\cite{ota2016stars} implemented a large scale parametrized RP framework (STaRS) with a heuristic optimization function. STaRS utilizes an efficient indexing scheme together with parallelism to simulate a variety of realistic scenarios. Alonso-Mora et al.~\cite{alonso2017demand} proposed an ILP-based dynamic RP framework by extending the concept of shareability graph~\cite{santi2014quantifying}, which reduces the dimensionality of the underlying assignment of ILP. Speed-up techniques for improving this model in terms of computational efficiency and system performance are proposed in~\cite{liu2019} and \cite{engelhardt2019}. Furthermore, a federated optimization architecture coupled with a one-to-one linear assignment problem in contrast to many-to-one assignment in~\cite{alonso2017demand} led to a more computationally efficient framework \cite{IBM}.

On-demand ride-hailing systems need to deal with the inherent inefficiency caused by demand asymmetry~\cite{Wei13,clemente2013vehicle}. Mobility-on-demand frameworks address this issue with the rebalancing strategies formulated as ILP models based on feedback proportional predictions~\cite{spieser2016shared, zhang2016control}, and by utilizing dynamic pricing~\cite{nosko2015, banerjee2015,dynamicRoutingRPMP}. In the ILP model implementation (i.e. our baseline), we employed a fast linear programming model minimizing the sum of travel times/distances incurred by idling vehicles to reach unserved requests.


Meeting points were first introduced in the context of RP to avoid detours, minimize the number of pickups and dropoffs, and increase the flexibility with additional assignments of requests to vehicles~\cite{stiglic2015meetingpoints}. Uber modified the idea of meeting points for an RP framework and came up with Xpress Pool service in which passengers may require to walk a few blocks to get picked up in exchange for a cheaper ride.  This also has led to the development of new queries in finding the optimal meeting point for a group of users~\cite{OMMR, MPP}. However, the limited number of studies that address the use of meeting points in RP are also constrained by the tradeoff of computational efficiency vs optimality~\cite{dynamicRoutingRPMP}. To the best of our knowledge, in literature, there is no RP framework with meeting points (or RPMP) that is both scalable and well performing in practice.

The backbone of an RP model or simulation framework is efficient shortest path queries. Shortest path\st{s} caching has been studied extensively to support fast querying in large scale road networks. The standard approach  to compute point-to-point shortest paths is Dijkstra's algorithm~\cite{Dijkstra1959} with a log-linear time complexity.
Variants of it cater to specific scenarios by using heuristics to improve the running time. However, in the context of RP services, running Dijkstra's algorithm or its variants on the fly is costly and unscalable when considering the size of modern day road networks. Highway Hierarchies~\cite{sanders2005hh}, reach~\cite{gutman2004reach}, and Contraction Hierarchies~\cite{geisberger2008ch} leverage the idea of shortcuts and the hierarchical structure of road networks to preprocess the graph to allow sub-linear time distance queries. Contraction Hierarchies in particular separates the preprocessing stage into two phases: (1) node ordering, and (2) hierarchy construction. Changing the weights of the graph only requires rerunning the hierarchy construction phase, thus making it a practical choice when real-time data such as road traffic conditions are involved. Transit Node Routing~\cite{Bast566} is a static caching method that reduces long-range queries to a number of table look ups. However, it suffers from slow local queries. Combining it with heuristic-directed search methods such as A* and Arc-Flags~\cite{hilger2009arcflags} can improve the overall performance. TNR+AF~\cite{bauer2008tnraf} in particular has been shown to be one of the fastest combinations. Hub-Labels~\cite{abraham2010a} provided a practical implementation of Hub-Labels and showed empirically that it is the fastest shortest paths caching scheme to date.
\section{ride-pooling model}
In this section we describe the components of our RP model in general and further elaborate on our baseline ILP-based optimization strategy.

\subsection{Model details}
The main components of our RP model are as follows:

\begin{figure*}[h]
    \begin{center}
    \includegraphics[width=0.8\linewidth]{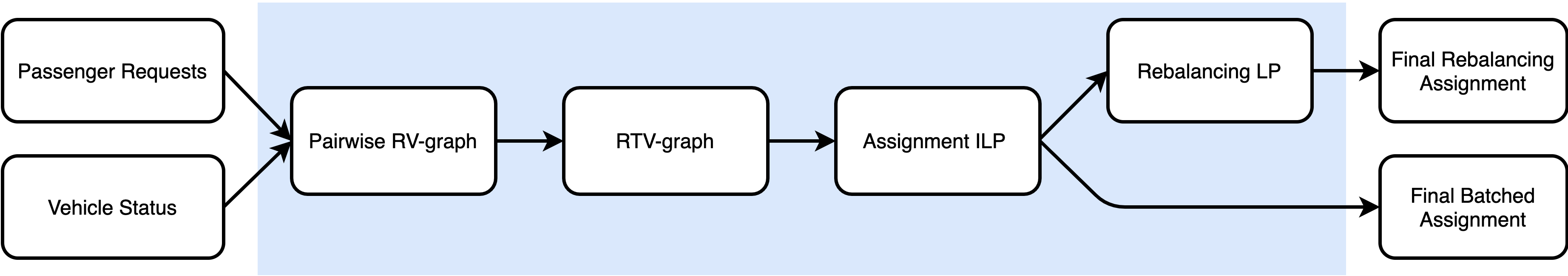}
    \caption{Schematic overview of the ILP-based methodology for batch assignment of multiple requests to multiple vehicles.} 
    \label{fig:FlowPNAS}
    \end{center}
\end{figure*}


\noindent\textbf{Road Network.} We define the underlying road network of the city as a directed weighted graph $G(\mathcal{V}, \mathcal{E})$ in which, each $v \in \mathcal{V}$ corresponds to an underlying road intersection and each $e \in \mathcal{E}$ corresponds to a road segment. The weight of each edge represents the length of the corresponding road segment. We use OpenStreetMap~\cite{OpenStreetMap} to extract the road network modeled as a graph. Note that two versions of the road network are used in this study. (1) The driving road network used for the purpose of vehicle movements is obtained as mentioned before while (2) the walking network used in ride pooling with meeting points is obtained by 
converting the driving network to an undirected graph. For each intersection $v$ in the walking network, the meeting point set $M_i$ is defined as the set of intersections which are in less than a maximum walking distance $D_w$ from $v$.

\noindent\textbf{Request Set.} A request is defined as a group of passengers associated with an origin, a destination, the request time, and the size of the group which is greater than or equal to one. Origin and destination in this formulation are the projections of the origin and destination derived from data to their closest intersection in the road network. The request set $R = \{r_1,...,r_n\}$ is the chronologically sorted collection of requests in which $r_i$ is a quadruple $({org}, {des}, {time}, {\#pass})$ representing the aforementioned four elements in order. 

\noindent\textbf{Vehicle Set.} Vehicle set refers to a fleet of $m$ vehicles $V= \{v_1,...,v_m\}$ that are used to serve the passenger demands throughout the simulation. Each $v \in V$ is a tuple $({id},{cap},{loc})$ in which ${id}$, ${cap}$ and ${loc}$ stand for vehicle identification number, capacity and initial location respectively.
In addition to the above specifications, each vehicle maintains a list of stops it has to traverse, its previous stop and the distance from previous stop alongside its current occupancy, speed and odometer reading.


\noindent\textbf{Constraints.} Parametrizing the simulation framework gives us the flexibility to incorporate different constraints that are imposed by the physical setup, performance or customer satisfaction. The following outlines the set of constraints (denoted by $Z$) used in the simulations. 
\begin{itemize}
    \item[] \textit{Precedence} : For each request $r \in R$, its origin $org$ should be visited before its destination $des$.
    
    \item[] \textit{Capacity} :  For each vehicle $v \in V$, at any given time, the occupancy can not be higher than the maximum \textit{Capacity} of the vehicle.
    
    \item[] \textit{Maximum waiting time} : For each request $r \in R$, waiting time $t_{wait}$  --- computed as the difference between pickup time and request time --- should not exceed the maximum waiting time $T_{wait}$ 
    
    \item[] \textit{Maximum travel delay} :  For each request $r \in R$, the amount of total travel delay $t_{total}$ (in vehicle delay plus waiting time) a passenger is willing to tolerate when sharing a vehicle with set of another passengers cannot exceed $T_{total}$. 

    \item[] \textit{Maximum Walking Distance} : In an RPMP service, passengers of a request $r \in R$ may be asked to walk to/from a certain meeting point given that it is within $D_w$ distance from the corresponding origin/destination.
\end{itemize}

\noindent\textbf{Scheduler.} This component is the core of the ride sharing model which encapsulates the generic functionality. Given the request set $(R)$, vehicle set $(V)$, road network $(G)$, and the constraints set ($Z$), it optimizes the model based on the optimization criteria 
and matches the requests to vehicles.

\noindent\textbf{Rebalancer.} The continuous non-deterministic flow of vehicles in a city together with the demand asymmetry may lead to poor functionality of an RP system. Rebalancing is the task of proactively relocating vehicles in the city to manage the demand with supply. The rebalancer is a component that imposes the rebalancing strategy on the fleet by assigning dummy destinations to idle vehicles. We further define and keep updating a rebalancing set denoted by $P$ as a set of candidate dummy destinations that are used for the purpose of rebalancing the vehicle fleet. It is important to note that we may reroute empty vehicles to serve actual requests while en-route to serve a rebalancer's request. In other words, we allow rebalancer's decision to be overridden by the scheduler decisions.

Parameterization empowers our model to simulate a wide range of scenarios and services with different constraints and optimization criteria. For instance, RP service can be simulated by setting the maximum walking distance $D_w$ to \textit{zero} while using a positive value for $D_w$ leads to RPMP. It is also worth mentioning that in an RPMP model, when either the vehicle or the passenger group who reaches the meeting point sooner than the other, they are expected to wait at the meeting point as long as it does not violate the constrains for either party.


\subsection{ILP-based optimization}

We first described the ILP-based optimization framework that is built upon the model and framework developed in~\cite{alonso2017demand} for real-time high-capacity RP. Our implementation makes some minor modifications to the RP model in~\cite{alonso2017demand} and extends their framework to enable RPMP, which is a new contribution of this work. 


The common framework is an iterative process that aggregates the demands over a fixed time interval and computes an allocation of vehicles to requests. Each iteration consists of four steps as depicted in Figure~\ref{fig:FlowPNAS}: computing a pairwise request-vehicle shareability graph, computing the graph of feasible trips and the vehicles that can serve them, solving an ILP to obtain the optimal assignment and finally a rebalancing step to re-distribute any remaining idle vehicles to align them with expected future demands. We provide a brief explanation of the aforementioned four steps below and refer the reader to~\cite{alonso2017demand} for further details. In Section~\ref{RPW_PNAS}, we provide a description of the extensions to the framework for incorporating meetings points. 

\subsubsection{Pairwise Request-Vehicle Shareability Graph (RV Graph)}
\label{RVGraph}
The shareability graph (\textit{RV Graph}), as defined in~\cite{santi2014quantifying}, is a graph construction that defines a node $i$ for each request $r_i$ and an edge between two nodes $(i,j)$ if the two corresponding requests $r_i, r_j$ can share the same vehicle (under some set of constraints that determine when requests can be shared). In our setting, two requests $r_i$ and $r_j$ are connected by an edge $e(r_i, r_j)$ if and only if a hypothetical vehicle starting at either of the two origins of the two requests can pick up and drop off both requests while satisfying the constraint set $Z$. Similarly, the \textit{RV Graph} also defines which vehicles can serve each of the requests. A vehicle $v$ and request $r$ is connected by an edge $e(v, r)$ if and only if vehicle $v$ can serve request $r$ without violating the constraint set $Z$. The weight of each edge $e(i,j)$ corresponds to the total time it takes for the two requests to be served (in R-R edges) and the times it takes for the vehicle to get to the origin of the request (in R-V edges). These travel times are computed using the $travel$ function as explained below.

Given a vehicle $v$ with a set of on-board passengers $P_v$ and a set of assigned requests $R_v$, we use the ${travel(v, R_v)}$ function to determine the optimal route $\sigma_v$ 
to pick up and drop off all the requests $P_v \cup R_v$, while minimizing the total travel time of the vehicle and satisfying the constraint set $Z$. If no valid route (one that does not violate $Z$) exists, the function returns "invalid". If there are valid routes, the function returns the total travel time of the vehicle according to the optimal route $\sigma_v$. This routing problem is a generalized Traveling Salesman Problem (TSP) that includes precedence constraints, time-windows and capacity, and computationally hard to even approximate. However, when vehicle capacities are small, the problem can be solved using an exhaustive search process with pruning. For settings with large vehicle capacities, heuristics can be used to reduce the computational overhead. A popular such heuristic is the insertion heuristic which only searches through the routes that preserve the order of the on-board passengers.


\subsubsection{Request-Trip-Vehicle Graph (RTV Graph)}
\label{RTV Graph}

Once the ${RV}$ ${Graph}$ is computed, the next step is to determine all feasible trips (grouping beyond single R-R and R-V pairs). A trip $T = \{r_1, .., r_t\}$ is a collection of $t$ requests that can be combined together and served by a single vehicle, while satisfying the constraints $Z$ for all requests in $T$. A trip is said to be feasible if it can be served by a vehicle while adhering to the constraints. Computing the feasible trips requires exploring the cliques (complete sub graphs) of the \textit{RV Graph}. Note that the existence of a clique is only a necessary condition (and not sufficient) for the feasibility of a trip that includes all the nodes in that clique. Each clique needs to be tested using the $travel$ function to makes sure that the trip is actually feasible~\cite{alonso2017demand}. Also note that a given request $r$ may belongs to more than one trip and each trip $T$ may be serviceable by more than one vehicle. 

The \textit{RTV Graph} contains two types of edges. A trip $T$ and a request $r$ are connected by an edge $e(r, T)$ if and only if $r \in T$. A trip $T$ and a vehicle $v$ are connected by an edge $e(T, v)$ if and only if $T$ can be served by vehicle $v$ without violating the constraints $Z$ corresponding to all $r \in T$. An edge weight of $d_{\sigma_v}$ as returned by the $travel$ function is associated with each edge $e(T, v)$. 

\subsubsection{Optimal Assignment (ILP)}
\label{ILP}

Given the set of all feasible trips $\mathcal{T}$ based on the \textit{RTV Graph}, we then compute the optimal assignment of trips to the vehicles. This optimal assignment is found via an ILP as formulated below.

For each edge $e(T_i, v_j)$ in the \textit{RTV Graph}, a binary variable $\epsilon_{i,j} \in \{0,1\}$ is created. We denote $\epsilon_{i,j} = 1$ if vehicle $v_j$ is assigned to trip $T_i$. Also, another binary variable $x_k \in \{0,1\}$ is introduced for each request $r_k \in R$. We denote $x_{k} = 1$ if request $r_k$ is not assigned to any of the vehicles in $V$. Let $\xi_{TV}$ be the set of $\{i, j\}$ indices for which an edge $e(T_i, v_j)$ exists in the \textit{RTV Graph} where as $\Gamma_j^{V}$ be the indexes $i$ for which an edge $e(T_i, v_j)$ exists in the \textit{RTV Graph}.

Then the cost function $C$ is defined as per Equation~\ref{Eq:ObjectiveFunction}.
\begin{equation}
C = {\sum_{i,j \in \xi_{TV}} c_{i,j}\epsilon_{i,j}} \quad  + \quad  {\sum_{k \in \{0,..,n\}} c_{ko} x_{k}}
\label{Eq:ObjectiveFunction}
\end{equation}

\noindent where $c_{i,j}$ denotes the $d_{\sigma_v}$ as returned by the $travel(v_j, T_i)$ and $c_{ko}$ denotes a large enough constant to penalize the ignored requests.

There are two constraints in the formulation, (1) each vehicle is assigned to at most one trip and (2) each request is either assigned to a vehicle or ignored. We define the two constraints as per Equation~\ref{Eq:ConstraintOne} and Equation~\ref{Eq:ConstraintTwo} respectively.

\begin{equation}
{\sum_{i \in {\Gamma}_j^{V}} \epsilon_{i,j}} \leq 1 \quad  \forall v_j \in V
\label{Eq:ConstraintOne}
\end{equation}
\begin{equation}
{\sum_{i \in {\Gamma}_k^{R}} \sum_{j \in {\Gamma}_i^{T}} \epsilon_{i,j} + x_k} = 1 \quad  \forall r_k \in R
\label{Eq:ConstraintTwo}
\end{equation}

\noindent where ${\Gamma}_k^{R}$ and ${\Gamma}_i^{T}$ denote the indices $i$ for which an edge $e(r_k, T_i)$ exists in the \textit{RTV Graph} and the indices $j$ for which an edge $e(T_i, v_j)$ exists in the \textit{RTV Graph} respectively.

\subsubsection{Rebalancing}
\label{Rebalance}

After each iteration of the batch assignment, we assign the remaining idle (empty) vehicles to unserved requests (if they exist) such that sum of the travel times is minimized. If there are fewer empty vehicles than there are unserved requests, all the vehicles will be assigned to subset of unserved requests. If there are fewer unserved requests than there are empty vehicles, some vehicles may remain idle, saving fuel and distance traveled. The problem is formulated as a standard rebalancing linear program~\cite{alonso2017demand, spieser2016shared} to determine the movement of vehicles.

\begin{figure*}[t]
\centering
\subfloat[]{
    \includegraphics[width=0.35\textwidth, keepaspectratio]{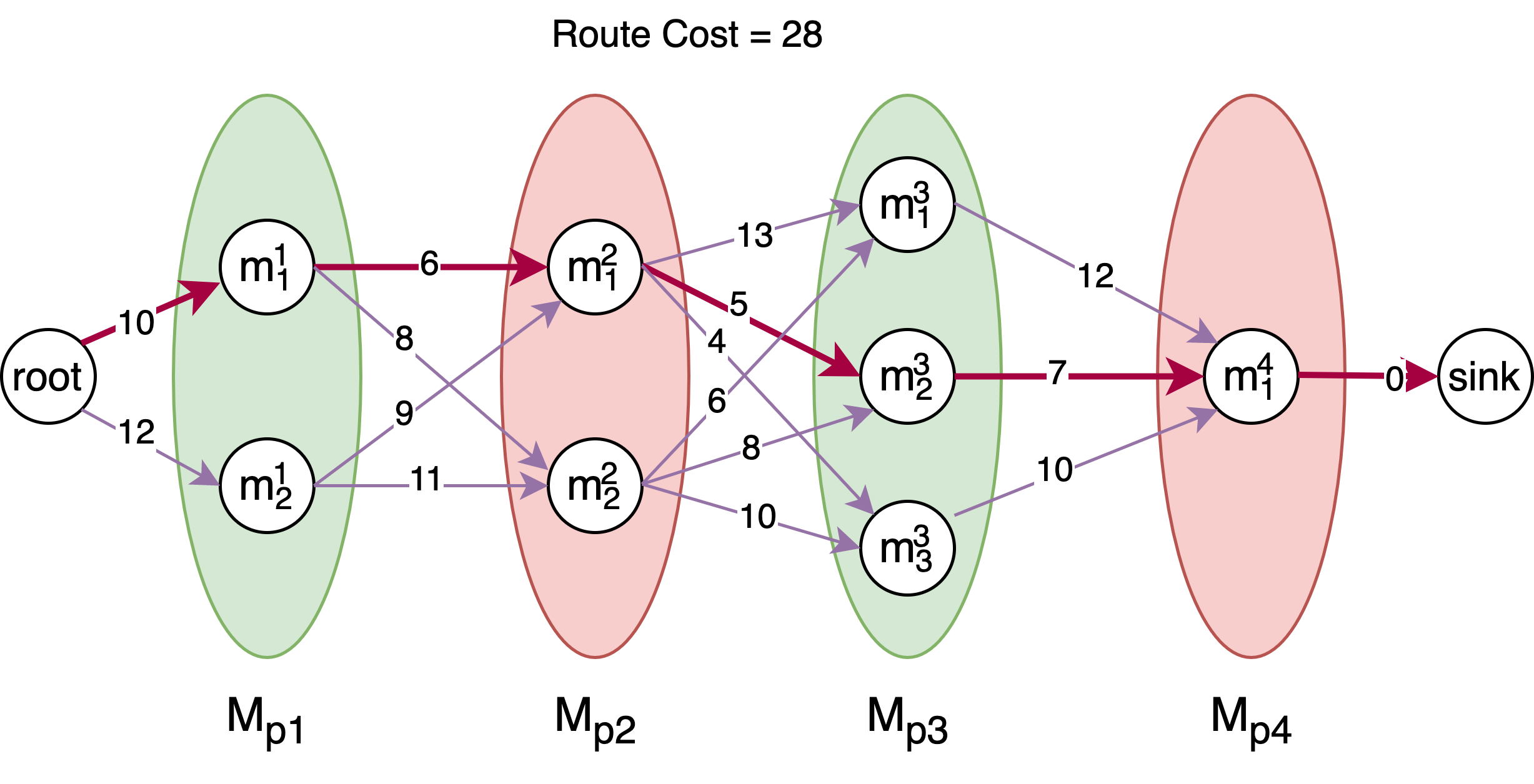}
}
\subfloat[]{
    \includegraphics[width=0.35\textwidth, keepaspectratio]{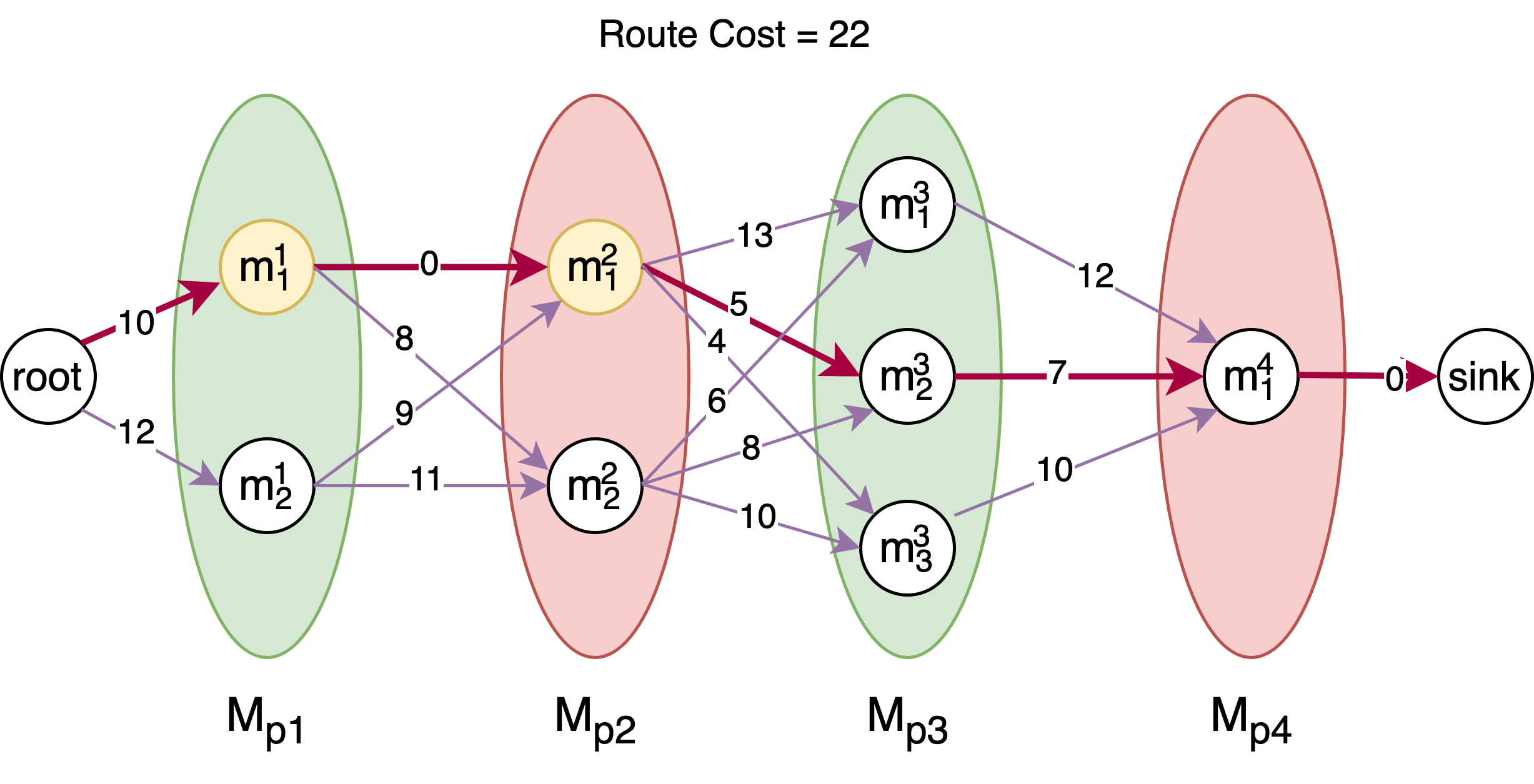}
}
\caption{An example instance of the directed acyclic graph $\tilde{G}$ for a vehicle $v$ and a route $\sigma_p$ with two new passengers. In $\tilde{G}$, both passengers are picked up before any dropoffs happen. Meeting point sets for the two passengers are denoted separately by the red and green partitions. (a) $\tilde{G}$ if the two passengers do not share meeting points. b) $\tilde{G}$ if the two passengers share a common pickup point and therefore the edge cost between $m_1^1$ and $m_1^2$ is zero.}
\label{fig:LevelGraph}
\end{figure*}


\subsubsection{ILP-based optimization with meeting points}
\label{RPW_PNAS}
In this section, we show how the framework from~\cite{alonso2017demand} can be extended to include RPMP. The only component of the workflow that needs to be modified is the computation of the vehicle routing in the $travel$ function. Even though the extension is isolated to this single component, including meeting points leads to significant additional complexity as will be evident in the following discussion.

In the $travel$ function for the 
RP problem, the generalized TSP needs to find the best route that serves the requests in a given trip $T$ while satisfying all the system constraints $Z$. In the RPMP problem, pickup/dropoff locations are not fixed and consists of a set of locations (nodes within the preferred walking distance $D_w$ of the pickup/dropoff location). This leads to solving a problem known as the group TSP, set TSP or covering TSP~\cite{coveringTSP, groupTSP} 
and related to the group Steiner Tree problem, which is significantly harder than TSP and is known to be at least as hard as the Set Cover problem, and 
challenging to even approximate~\cite{garg2000polylogarithmic}.  

Our approach for solving this problem consists of two steps. We first define a set of routes, where each $route$ corresponds to a fixed ordering of the pick ups and dropoffs of the trip $T$. Let $\mathcal{R}$ = $\{\sigma_1,...\sigma_n\}$ be the set of all feasible routes for the vehicle $v$ with the set of on-board passengers $P_v$ and the set of assigned requests $R_v$. The feasible route set is obtained by removing routes that violate the precedence and vehicle capacity constraints. Let $v_{current}$ be the current location of the vehicle $v$ and 
$X_i =\{x_1,...,x_a\}$ be the set of pickup/dropoff points of the route $\sigma_i$. Each $x_j \in X_i$ has a set of meeting points defined as $M_{ij} = \{m_1^j, ...m_b^j\}$.


Next, for each valid route $\sigma_p \in \mathcal{R}$, we build a directed acyclic graph (DAG) ${\tilde{G}} = (\tilde{\mathcal{N}}, \tilde{\mathcal{E}}, \tilde{\mathcal{W}})$ such that the nodes set $\tilde{\mathcal{N}}$ consist of all the meetings points in $M_{pj}$ for all $x_j \in X_{p}$, $v_{current}$, and a sink node $s$. The graph is built as follows: (1) Add the vehicle current location $v_{current}$ as
$root$
; (2) For each pickup/dropoff point
$x_j \in X_{p}$, create nodes for the set of corresponding meetings points $M_{pj}$; and (3) Add a sink node $s$. Since the order of the route is fixed, the edges $\tilde{\mathcal{E}}$
only consist of the following pairings: (1) two nodes $m_u^i \in M_{pi}$ and $m_v^j \in M_{pj}$ are connected when $i=j+1$; (2) the vehicle node $v_{current}$ is connected to the all the meeting points in $M_{p1}$; and (3) the sink node $s$ is connected to the last dropoff points in $M_{pa}$. This defines three types of edges in the graph. The edge weights set $\tilde{\mathcal{W}}$ is formed by weights of the edges as given by the travel time between the corresponding nodes except for the weights to the sink node which are all zero. An example instance of $\tilde{G}$ is illustrated in Figure~\ref{fig:LevelGraph}.

For each route,  we run Algorithm~\ref{alg:PNAS-XPOOL_levelGraph} on the corresponding graph $\tilde{G}$ to obtain the total cost of the route. Algorithm~\ref{alg:PNAS-XPOOL_levelGraph} returns the total travel time required to traverse all the pickups/dropoffs of trip $T$ while ensuring that the maximum walking distance, maximum waiting time, and maximum travel delay constraints are not violated for any of the new requests $R_v$ or on-board passengers $P_v$. If such a route does not exist, the algorithm will terminate and state so. When the capacity of the vehicle is fixed, each pickup/dropoff point has $\mathcal{O}(k)$ meeting points, and the trip has $|T|$ requests, the complexity of this algorithm is $\mathcal{O}(k^2|T|)$. 


Let $D = \{d_{\sigma_1},..,d_{\sigma_n} \}$ be the set of shortest distances as returned by the Algorithm~\ref{alg:PNAS-XPOOL_levelGraph} for all routes in $\mathcal{R}$. Finally we take the minimum of $D$ as the minimum distance and return as the response to the $travel$ function.

\begin{algorithm}[t]
 \caption{FindOptimalRoute ($G^\prime$, root, sink)}
 \label{alg:PNAS-XPOOL_levelGraph}
 \SetAlgoLined
 \KwData{$G^\prime$ = ($N^\prime$, $E^\prime$), root, sink}
 \KwResult{Optimal path cost if exists or "Invalid" otherwise }
 \ForEach{$v \in N^\prime$ }{
    dist[v] := $\infty$ \\
    previous[v] := undefined \\
 }
 dist[root] $\gets$ 0 \\
 $Q \gets \emptyset$ \\
 $Q.push$ ($root$, 0)\\
 \While{$Q$ $\not $= $\emptyset$}{
    $(u, cost) \gets Q.pop()$ \\
    \ForEach{$e(u,v) \in {E^\prime}$}{%
        $d_{root \rightarrow v}$ =  $cost$ + $w_{u,v}$ \\
        ${d^\prime_{root \rightarrow v}}, status \gets$ checkConstraints($v, d_{root \rightarrow v}$) \;
        \If{\text{status is valid} \textbf{and} ${d^\prime_{root \rightarrow v}} \textless dist[v]$ }{
            \eIf{$Q$ contains $(v, dist[v])$ }{
                $Q.update((v, {d^\prime_{root \rightarrow v}}))$ \\
            }{
                $Q.push((v, {d^\prime_{root \rightarrow v}}))$ \\
            }
            $dist[v] \gets {d^\prime_{root \rightarrow v}}$ \\
            $previous[v] \gets u$\
        }
    }
 }
 \If{$dist[sink] \; is \; set$ }{
    \Return $dist[sink]$
 }
\Return invalid
\end{algorithm}

%
We use a specialized shortest path algorithm, shown in Algorithm 1, to compute the optimal path from $root$ to $sink$ in $\tilde{G}$. Algorithm~\ref{alg:PNAS-XPOOL_levelGraph} maintains a priority queue $Q$ where each element in $Q$ is a tuple of the form $(v,cost)$. Here, $v \in \tilde{\mathcal{N}}$ and \textit{cost} is the path cost from $root$ to $v$, which also acts as the priority factor. Our queue has three main operations, \textit{push, pop} and \textit{update}. The $push$ operation is used to enqueue elements to the queue. The $pop$ operation dequeues the element with the lowest $cost$ value whereas $update$ operation is used to update an element in the queue. Apart from that we also define two maps $dist$ and $previous$. $previous$ 
stores the predecessor for each $v \in \tilde{\mathcal{N}}$ in the shortest path to $v$. Similarly, we use $dist$ to store the shortest travel time to reach each $v \in \tilde{\mathcal{N}}$ without violating constraints. We further define the weight of the edge $e(u,v)$ as $w_{u,v}$ and the function \textit{checkConstraints} which takes a node $v$ and the cost of the path up to node $v$ as inputs. It returns two outputs, (1) status : whether the current path cost does not violate any of the imposed constraints and (2) the modified path cost to accommodate vehicle waiting time at the node $v$ for any passenger who takes more time to reach to the node. For example, if the cost of the path so far to the node $v$ is $c$ and the vehicle has to wait for $t_{hold}$ 
at node $v$ for a given passenger to walk to the node, \textit{checkConstraints} returns $\bar{c}$ as the new modified path cost where $\bar{c} = c+t$. Note that if the vehicle does not have to wait for the passenger, then $t=0$ and thus $\bar{c} = c$. This setting captures the situations where the vehicle reaching a given meeting point before the passenger does and other way around.

The algorithm first initializes \textit{Q, dist}, and \textit{previous} (line 1-6). Next, it adds the $root$ node to $Q$ with a path cost of zero length (line 7). Then while $Q$ is not empty, we \textit{pop} elements from $Q$ and expand from each element $(u,cost)$ by visiting its adjacent vertices. For each
new exploration of adjacent node $v$, we check the constraints (line 12). If they are satisfied and the modified cost $\bar{c}$ is less than the current shortest travel time cost estimate to $v$ as given by $dist[v]$, we update $Q$ and the two maps (line 13-21). Finally, once $Q$ is empty, we return the value of $dist[sink]$ or "invalid" if no valid path was found (line 24-28).

\subsubsection{Correctness of Algorithm 1}
\label{correctness}

We show the correctness of the the Algorithm~\ref{alg:PNAS-XPOOL_levelGraph} by contradiction. Given the directed acyclic graph $\tilde{G} = (\tilde{\mathcal{N}}, \tilde{\mathcal{E}}, \tilde{\mathcal{W}})$, let $d(root, x)$ be the minimum travel time between $root$ node and any node $x \in \tilde{\mathcal{N}}$ such that all constraints are satisfied over the path to $x$. Note that $d(root, x)$ is flexible to contain the time the vehicle spends waiting for passengers at meeting points. Below we show that the travel time from $root$ to any node $x$ as returned by the Algorithm~\ref{alg:PNAS-XPOOL_levelGraph}, $dist[x]$, is equal to the minimum travel time $d(root, x)$. 

In Algorithm~\ref{alg:PNAS-XPOOL_levelGraph}, when a node $x$ is popped from the queue, we have that $dist[x] = d(root,x)$. Assume that this observation is false and there are some nodes such that when a node $x$ is popped from the queue, $dist[x] > d(root,x)$. Let $x$ be the first such node to violates the observation. This means all the nodes $l$ that were popped before $x$ adhere to the property $dist[l] = d(root,l)$ and only node $x$ demonstrates $dist[x] > d(root,x)$. Let $\sigma^*_p$ be the path that adheres to the constraints and minimizes travel time from $root$ to $x$ for a given route $\sigma_p$. When node $x$ is popped, let $z$ be the first node in $\sigma^*_p$ that is not yet popped whereas $y$ being the predecessor node of $z$ in $\sigma^*_p$ that has already been popped. We can conclude $dist[y] = d(root, y)$ and $dist[z] = d(root,y) + w_{yz} + t_z$ where $t_z$ denotes the waiting time at node $z$ if the vehicle has to wait for passengers. When the node $x$ is popped by the Algorithm~\ref{alg:PNAS-XPOOL_levelGraph}, we have, $dist[x] \leq dist[z]$. According to the principle of optimality~\cite{bellman}, a sub path of a shortest path is itself a shortest path. Therefore $d(root,x) = d(root,z) + d(z,x)$. Since we already know that, $dist[x] \leq dist[z]$, we can conclude $dist[x] \leq dist[z] = d(root,y) + w_{yz} + t_z \leq d(root,y) + w_{yz} + t_z + d(z,x) = d(root, x)$. This gives us $dist[x] \leq d(root,x)$ which contradicts with our initial assumption $dist[x] > d(root,x)$. Therefore we have that $dist[x] = d(root,x)$ for all $x \in \tilde{\mathcal{N}}$.


\subsubsection{Heuristics for real-time execution}
\label{heuristics}

For real-time execution, we could employ heuristics to speed up the process pipeline. Timeouts can be set for exploring candidate trips for vehicles in \textit{RTV graph} and solving the assignment ILP and rebalancing LP. It is also effective to limit the number of \textit{RV} edges of the \textit{RV graph}. In particular, the complete \textit{RV graph} can be pruned to contain the $RV$ edges corresponding to first $n$ vehicles with lowest service costs for each request.

In the computation of the RV graph with meeting points, one may first compute the RV edges without considering the meeting points
to reduce the additional overhead induced by the meeting points on the \textit{travel} function. One could then relax the waiting time and travel delay constraints by a specified constraint flexibility factor to identify the $RV$ edges that could potentially be viable if we use meeting points in the first place. By using the insertion heuristic to check the route viability for those identified potential $RV$ edges, one could 
add them back to the \textit{RV Graph} with significantly less overhead in real-time execution. 

It should also be noted that we could parallelize the computation of the \textit{RV Graph} and \textit{RTV Graph}. Most standard ILP solvers implicitly use parallelized pipelines to solve large-scale optimization problems. 
The heuristics and their configurations that are employed in our simulations are discussed in Section~\ref{accuracy_eval}.

%
\begin{figure*}[htp]
    \parbox[b]{.3\textwidth}{
        \subfloat[]{%
            \includegraphics[width=0.5\textwidth,keepaspectratio,height=4.5cm]{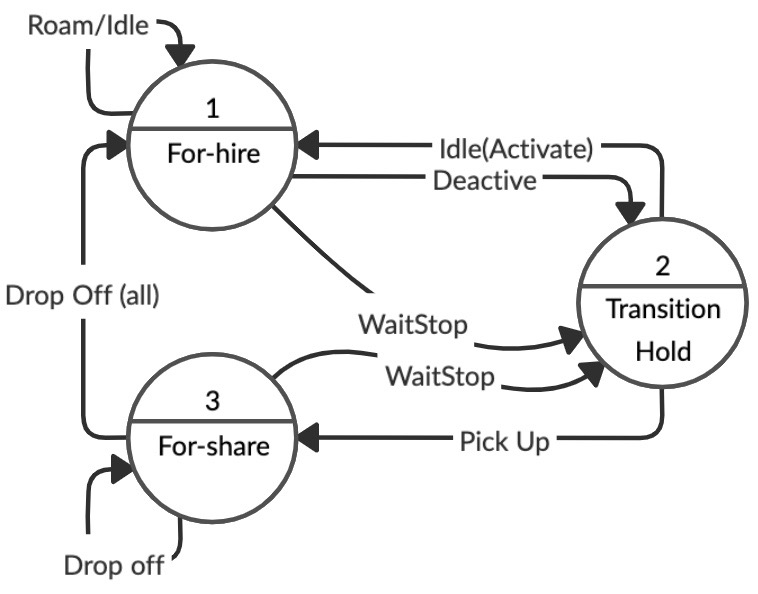}
            \label{fig:state machine}
    }
    \parbox[b][140pt][t]{100pt}{
        
        \subfloat[]{
               
            \resizebox{170pt}{!}{
                
                \begin{tabular}[b]{ll}
                Input Name & Description\\
                \hline
                Drop Off & $[s_{wait} = 0,\ s_{\#pass} <0,\ s_{distance}\geq 0]$  \\
                Pick Up  & $[s_{wait} = 0,\ s_{\#pass} >0,\ s_{distance}\geq 0]$ \\
                Wait Stop & $[s_{wait} \geq 0,\ s_{\#pass}=0,\ s_{distance}\geq 0]$ \\
                Deactivate & $[s_{wait} > 0,\ s_{\#pass} =0,\ s_{distance}=0]$ \\
                Activate & $[s_{wait} = 0,\ s_{\#pass} =0,\ s_{distance}=0]$\\ 
                Roam & $[s_{wait} = 0,\ s_{\#pass} =0,\ s_{distance}>0]$ \\
                Idle & $[s_{wait} = 0,\ s_{\#pass} =0,\ s_{distance}=0]$ \\
                \end{tabular}
                \label{table:stops}
            }
        }
        \\
        \subfloat[]{
            \resizebox{310pt}{!}{
                \begin{tabular}{c|ccccccc}
                \backslashbox{Current State}{Input Stop} 
                & Idle & Roam & Deactivate & WaitStop & Pickup & Dropoff \\
                \hline
                For-hire (1) & 1 & 1  & 2 & 2 & - & -\\
                Transition-hold (2) & 1 & - & - & - & 3 & -\\
                For-share(3) & - & -  & - & 2 & 3 & 3\\
                \end{tabular}
                \label{transition table}
            }
        } 
    }
  }
  \caption{The state machine (a), input table (b) and the transition table (c) of vehicle object in STaRS+.}
  \label{}
\end{figure*}

\section{Ride-pooling at scale}

The computational complexity of the ILP-based model for RPMP (see Section~\ref{RPW_PNAS}) requires extensive computational resources to be executed at scale. This is primarily due to the complexity induced by the \textit{travel} function in computing optimal routes for the fleet of taxis. 
%
%
To address the limitation of the ILP-based model, we introduce STaRS+ with a novel mechanism for caching shortest paths in large graphs, and an efficient heuristic for scheduling RPMP services at scale.
In the following sections, we elaborate on how we modified and extended STaRS~\cite{ota2016stars} and transformed it into STaRS+.

\subsection{STaRS+ Optimization Strategy }
\label{stars_optimization}
STaRS~\cite{ota2016stars} optimization strategy is based on a heuristic algorithm that optimizes the total cost using the minimal serving cost at the time of request.
%
%
In particular, the accumulated cost $T_i$ after receiving \emph{i} requests can be computed from $T_{i-1}$ and the minimum cost of serving the \emph{i}-th request $r_i$ considering every vehicle $v_j \in V$, denoted as $\mathcal{C}(r_i, v_j)$:
  \[
    T_i=\left\{
                \begin{array}{ll}
                    \begin{tabular}{ l }
                    $T_{i-1}$ + $min_{0\leq j \leq |V|}{\mathcal{C}(r_i, v_j)}$  \\
                    $T_{i-1}$  , if $r_i$ cannot be served\\  
                    \end{tabular}
                \end{array}
              \right.
  \]

As such, for each request $r_i \in R$, the locations of all vehicles in the fleet are first updated as part of their assigned itinerary or due to the rebalancing process that redistributes idle vehicles throughout the city. Next, the heuristic insertion process
%
%
computes the cost of serving $r_i$ for each available vehicle. Finally, $r_i$ is added to the stop list of vehicle $v_j$, where $r_i$ could be served with the minimum imposed cost. If it is not possible to serve a request without violating the simulation constraints, STaRS would add the requested origin to the \emph{rebalancing} set $P$. In the following, we describe our improvements on the STaRS model as well as how the aforementioned steps are transformed in STaRS+.

\noindent \textbf{Model Imporvement.}
A vehicle in STaRS is always moving either for seeking new passengers or for serving existing ones. Though these two states are sufficient to model most RP services, additional states are needed for RPMP. Specifically, we need to be able to express cases where a vehicle gets idle, or waits for a passenger, or becomes unavailable between shifts. In STaRS+, we propose the use of a state machine in which vehicles may transition among three different states named \textit{For-hire}, \textit{For-share}, and \textit{Transition-hold} as shown in Figure \ref{fig:state machine}. \textit{For-hire} indicates that a vehicle is idle or roaming without any passenger; \textit{For-share} indicates that a vehicle has at least one passenger on board; and \textit{Transition-hold} indicates that a vehicle has to wait for passengers at a pickup point or its needs to be on hold for a period of time.

STaRS+ also extended the notion of \emph{stop list} in STaRS, which maintains a list of stops that each vehicle has to pass by during the simulation. Not only that this stop list helps STaRS+ to regulate the pickup and dropoff order of a request, it also makes sure vehicles are relocated to locations in demand for rebalancing purposes.
For each served request, we add three types of stops to the stop list of the assigned vehicle: a \textit{Wait Stop} to instruct the vehicle to reach the pickup point, a \textit{Pick Up} stop to instruct the vehicle to wait for $t_{hold}$ seconds for the passenger, and a \textit{Drop Off} stop to instruct the vehicle to get to the requested dropoff point.
For vehicles that do not have any passengers, they will be appointed with one of the following three stop types: \emph{Idle}, \emph{Roam}, or \emph{Deactivate}, to indicate whether they are idling, roaming, or unavailable, respectively.

%
Each stop is further described by three variables: $s_{wait}$ -- the amount of time vehicle needs to wait at the stop, $s_{pass}$ -- the number of passengers to be picked up or dropped off, and $s_{distance}$ -- the distance from the previous stop to this stop. The descriptions of these stops are illustrated in Figure \ref{table:stops}, where stops are inputs of the state machine. 
The output state of the machine given each stop is presented in Figure \ref{transition table}. 
Note that the \emph{Deactivate} stop can be used to model scenarios where vehicles are unavailable, thus, allowing a fleet to take excessive vehicles offline as needed, e.g. during off peak hours.

\subsubsection{State Update}
For request $r_i$ issued at time $t$, we update the state of all vehicles in the vehicle set $V$ and identify their current position based on the elapsed time since the last update. Next, stops that were visited by a vehicle during this period, e.g. as a result of pickup or dropoff, will be removed from the vehicle's stop list. In the case of roaming vehicles, for those that reached their rebalancing destination, they will become idle.
Otherwise, if a vehicle has not reached its target, its current position will be computed as the closest intersection along the traveling route.
\subsubsection{Rebalancing}
Once a vehicle gets idle during the update process, the rebalancing process is triggered.  This process searches the rebalancing set $P$ and picks the closest destination to the vehicle's current location as its next destination. If $P$ is empty, the vehicle is kept idle at the location of its last stop.

Though rebalancing improves the efficiency of the system, poor strategies may worsen the traffic problem by increasing the rate of empty vehicles driving in the city. STaRS sets $P$ to be all intersections of the road network, and randomly choose a destination from $P$ each time a vehicle needs to be redistributed. The first two rows of Table \ref{rebalancing} shows the performance of the system without and with STaRS's rebalancing strategy. Comparing them, we can admit the huge improvement in the service rate which comes at the expense of 
higher average vehicles miles driven (this parameter is described in Table~\ref{Notations}). 
Consequently, in STaRS+, $P$ is set to only the locations of unserved requests within a batching time window $\Delta$. Idle vehicles are then redistributed to the closest spots among these locations. 


The rationale behind the rebalancing strategy of STaRS+ is that unserved requests usually occur when there is a spike in service demand (e.g. when there is an event happening) or due to the imbalanced distribution of vehicles. 
%
In either case, the origins of unserved requests can be considered as a proxy to the potential future demands, thus, rebalancing vehicles to these locations would likely improve the overall serving rate.
As illustrated in Table~\ref{rebalancing}, the serving rate using the strategy in STaRS+ is observed to be significantly higher than without any rebalancing efforts.
Though not outperforming the strategy in STaRS, this approach practically alleviates the undesirable effect of randomly rebalancing vehicles 
without sacrificing performance. 


\subsubsection{Sharing Cost Estimation}
The cost function $\mathcal{C}(r_i, v_j)$ calculates the minimum cost of adding request $r_j$ to the stop list $S_{v_j}$ of vehicle $v_j$. This cost is the weighted sum of the extra time (as a proxy of distance) that vehicle $v_j$ needs to travel to accommodate $r_i$ in its trip plan $\delta_{(r_i, S_{v_j})}$ and the probable waiting time at the pick-up spot in RPMP service $t_{\text{hold}(v_j, r_i)}$.

  $$\mathcal{C}(r_i, v_j) = c_1 . \delta_{(r_i, S_{v_j})} + c_2 . t_{hold}(v_j, r_i)$$
  
The cost function can be modified to fit the optimization criteria. In our setting, the cost function reflects the additional time needed for the vehicle to serve a specific request. As a result, the weight of extra time $c_1$ is equal to the weight of waiting time $c_2$. However, in the case of optimization for fuel consumption, for example, $c_1$ could be set higher than $c_2$ to reflect the importance of the distance driven over $t_{hold}$.
\begin{table}
    \centering
    \resizebox{210pt}{!}{
    \begin{tabular}{c c c}
       \hline
        Strategy & RP & RPMP \\
        \hline
        None &  \begin{tabular}{c}
                Service Rate:  37.00\% \\
                VMT:  156.31\\
            \end{tabular} &
            \begin{tabular}{c}
                Service Rate:  57.05\% \\
                VMT:  162.26\\
            \end{tabular}\\
        \hline
        STaRS & \begin{tabular}{c}
                Service Rate:  \textbf{82.29}\% \\
                VMT:  370.50\\
            \end{tabular} &
            \begin{tabular}{c}
                Service Rate:  \textbf{94.78}\% \\
                VMT:  238.24\\
            \end{tabular}\\
        \hline
        STaRS+ & \begin{tabular}{c}
                Service Rate:  65.91\% \\
                VMT:  260.04\\
            \end{tabular} &
            \begin{tabular}{c}
                Service Rate:  81.79\% \\
                VMT:  221.25\\
            \end{tabular}\\
        \hline
    \end{tabular}
    }
    \caption{The performance of the system without and with rebalancing methods used in STaRS and STaRS+. VMT represents average miles driven by vehicles during simulation 
    .}
    \label{rebalancing}
\end{table}
The problem of finding the optimal stop order (to minimize the cost function) is a variation of the traveling salesman problem known as \emph{asymmetric traveling salesman problem with precedence constraints}~\cite{ATSP}. In our case, each vehicle maintains a list of stops with precedence constraints such as a request pickup must appear before its corresponding dropoff. STaRS suggests a heuristic for this NP-hard problem. The heuristic tries to find the best pair of indices for inserting the pickup and dropoff of request $r_i$ into vehicle $v_j$'s stop list $S_{v_j}$. Assuming the dropoff has been added to the end of the stop list, the algorithm finds the index for which adding the pickup stop imposes the minimal extra cost. To prune the search space for the optimal index, STaRS finds the minimum index (in the stop list) after which the vehicle's capacity allows accommodation of the new request. Next, it searches for the optimal pickup and dropoff index while maintaining the global simulation constraints. As such, a passenger must be picked up no later than $T_{wait}$ time, with no more than $T_{extra}$ detours incurred, and the total added time to his or her trip must not exceed $T_{total}$.
Note that for RPMP, $t_{hold}$ may change along the route of a passenger due to others joining his or her ride.

\subsubsection{Ride-Pooling with Meeting Points}
Thanks to the flexibility of the sharing cost model, STaRS can be extended to support ride-pooling with meeting points (RPMP) without major changes to the simulation process. This extended version is called STaRS+.

Let $r=\langle x_p,x_d \rangle$ denote a request with $x_p$ and $x_d$ as its pickup and dropoff location. The corresponding set of meeting points for $x_p$ and $x_d$ are $M_p$ and $M_d$, respectively. Unlike RP, where we optimize the sharing cost 
of a single pair of stops ($r$), in the case of RPMP, we need to optimize the sharing cost across all possible pairs of meeting points. In other words, our objective is to find the meeting points $m_p$ and $m_d$ such that the cost of assigning $\langle m_p,m_d \rangle$ to vehicle $v$, or $\mathcal{C}(\langle m_p,m_d \rangle,v)$, is minimum $\forall m_p \in M_p$ and $\forall m_d \in M_d$. 
A brute-force approach that tries all possible pairs has a time complexity of $\mathcal{O}(|M_p||M_d|)$, and is not suited for large RPMP simulations. This is because the optimization must be done for every request and every vehicle, yielding the total time complexity of $\mathcal{O}(|M_p||M_d||V||R|)$, where $|V|$ and $|R|$ represents the size of the vehicle set and the request set, respectively. To reduce the complexity, STaRS+ relies on an iterative optimization heuristic to determine $m_p$ and $m_d$. First, we fix $x_d$ as the best candidate for $m_d$, and find an $m_p \in M_p$ that minimizes $\mathcal{C}(\langle m_p,x_d \rangle,v)$. Then with $m_p$ as the optimal pickup point, we find an $m_d \in M_d$ that minimizes $\mathcal{C}(\langle m_p,m_d \rangle,v)$. This two-stage approach has the time complexity of  $\mathcal{O}((|M_p|+|M_d|)|V||R|)$.

Nevertheless, the number of optimizations could grow significantly large as the number of meeting points increases with a higher preferred walking distance, particularly for a dense road network like NYC. To reduce the search space, STaRS+ utilizes the minimum waiting time constraint to prune vehicles that are not able to reach any of the pickup meeting points in time.
For the pruning process, we define an upper bound radius using the great-circle distance between the current location of the vehicle and the request's origin, i.e vehicles that are outside of this radius are not eligible to serve the request. The upper bound is the sum of the preferred walking distance ($D_w$) and the great-circle distance equivalent of the maximum waiting time ($T_{wait}$) of the passenger. Intuitively, it considers the best-case scenario in which both the passenger and the vehicle move toward each other on the direct line. This pruning strategy effectively reduces our search space by over 80\% and leads to a 5x speedup for supporting RPMP in STaRS+.

\subsubsection{Batch Assignment}
In STaRS+, we employ a similar strategy for batch assignment as described in Section ~\ref{ILP}. First, requests and potential matching vehicles are grouped into batching windows of $\Delta$ time. Then, we assign vehicles to requests that optimizes a cost function. The batch assignment problem can be formulated as a bipartite graph matching problem, where both the unweighted version (USTaRS+) and the weighted version (WSTaRS+) are considered. By default, the cost function in USTaRS+ and WSTaRS+ are the number of matches and the vehicle miles driven, respectively. However, the cost function in WSTaRS+ can be redefined by users depending on the scenario. With respect to time complexity, USTaRS+ optimizes the matches in each batch in $\mathcal{O}(|V|^2)$ time, 
while WSTaRS+ implements the Hungarian algorithm~\cite{Hungarian} that can minimize the total cost in $\mathcal{O}(|V|^3)$ time.
To reduce the complexity of many-to-one matching, we frame the batch assignment in STaRS+ as a linear assignment problem defined in~\cite{IBM}. In linear assignment problem at most one request is assigned to each vehicle in each batch (one-to-one matching). Results of experiments represented in section~\ref{accuracy_eval} show when the size of the batch is small enough, the one-to-one matching strategy could perform almost as good as many-to-one in ride-pooling with meeting-point.

\subsection{Shortest Paths Caching}\label{caching_scheme}

Shortest distances and shortest paths are frequently queried during simulation and contribute to a significant portion of the execution time. In the previous work~\cite{ota2016stars} where the geographical scope was only limited to the borough of Manhattan, storing the full distance and predecessor matrices was an acceptable solution. However, as the proposing framework aims to support graphs as large as all of NYC, storing the full matrices would require too much memory. Contraction Hierarchies can accommodate distance queries through a fast bidirectional Dijkstra search and accommodate path queries through an unpacking procedure. However, it is slow compared to static cache methods. Hub-Labels is the fastest amongst such methods that generalize the distance matrix. However, it is still relatively slow compared to direct cache look up and would cause the running time to increase by a significant factor. In order to support simulations over the five boroughs of NYC, or possibly larger graphs, a new caching scheme is required to achieve better scalability in terms of memory usage and running time. We devised a method that sacrifices minimal query time to achieve substantially less memory usage at the cost of preprocessing time. 

\subsubsection{Shortest distance Queries} 
Similar to the principle behind existing caching schemes, our approach to storing the shortest distances exploits the idea that there is a small selection of nodes that can cover all shortest paths. However, rather than finding a selection that covers all shortest paths, our method is focused on finding multiple selections for subsets of the shortest paths. In doing so, we were able to better leverage the notion of access nodes, and increase the compactness of the cache while preserving pseudo-constant-time distance queries. 

The idea is that when traveling on a road network from one intersection to any intersection in a different region via shortest paths, one will most likely take a common path before diverging to reach the respective destinations in that region. From our experiences with road networks, it is easy to see that there will likely be a longer common path when the starting intersection is farther away from the destinations. Our algorithm exploits this common sub-path property to create a compact cache structure.

\paragraph{Definition}
Let $G(V, E)$ be the road network, where $V$ is the set of intersections and $E$ is the set of streets. A distance query $dist(u, v)$ is the shortest distance from $u$ to $v$, where $u, v \in V$. A path query $path(u, v)$ is the list of all nodes that appear on the shortest path from $u$ to $v$ in the order they are visited. $dist(u, v)$ can also be answered through $dist(u, w)+dist(w, v)$ if $w \in path(u, v)$. A path $p$ is a common sub-path of a set of shortest paths $SP$ if it satisfies $p \subseteq sp$, for every shortest path $sp \in SP$. Such a path $p$ always exists if $SP$ is a set of shortest paths that start from a common source $u$. This property is guaranteed by the trivial common sub-path $p = \{u\}$.

Let $u \in V$ be the source, $D \subseteq V$ be the set of destinations, $SP_u$ be the set of all shortest paths with source $u$ and destinations $v \in D$, and $p_u$ be the common sub-path of $SP_u$. For any node $w \in p_u$, all distance queries $dist(u, v), v \in D$, can be expressed as $dist(u, w)+dist(w, v)$ since $p_u$ is a common sub-path. In other words, there is no need to cache the shortest distances if $dist(w, v), v \in D$ is already cached. $dist(u, w)$ is the only new information needed to accommodate the new queries. We say that $u$ is covered by $w$ in this case, and the node that covers $u$ is denoted by $cover(u)$.

\subsubsection{Hitting-Set}
The goal of the cache is to reduce memory usage and therefore we require a minimal set of nodes $C$ such that $cover(u) \in C$ for all of $u \in V$. This problem can be formulated as the Hitting-Set problem by using all of $p_u, u \in V$ as sets. The Hitting-Set problem is known to be NP-Hard, along with its equivalence that is more commonly studied, the Set-Cover problem. The greedy estimate adds the node that covers the most uncovered paths to the hitting set at each iteration, which is equivalent to choosing the set with the largest cover in Set-Cover \cite{johnson1973setcover}. It is known that the greedy heuristic for Set-Cover provides a $ln(n)$-Approximation (or is at most $ln(n)$ times larger than the optimal set), where n is the size of the largest set~\cite{chvatal1979setcover_bounds}. 

\begin{figure}[t]
\centering
\subfloat[Road Network]{
    \includegraphics[width=0.26\textwidth, keepaspectratio]{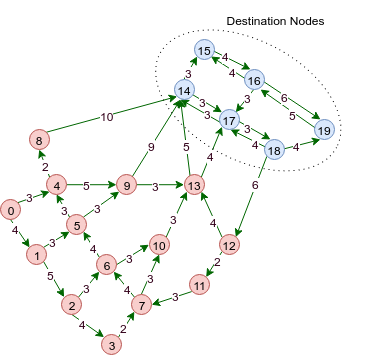}
}
\\
\subfloat[Cache partition for $D=\{14, 15, 16, 17, 18, 19\}$]{
    \includegraphics[width=0.38\textwidth, keepaspectratio]{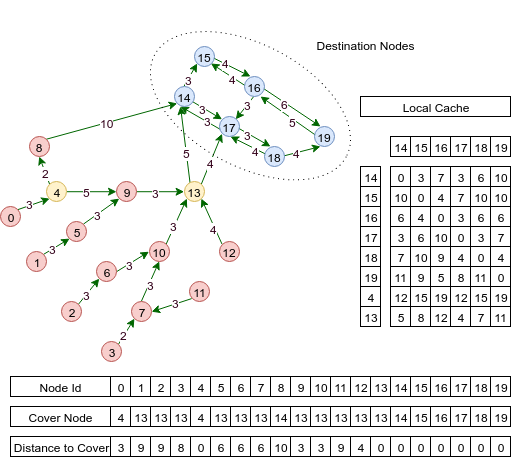}
}
\caption{Example of a cache partition: (a) a road network with 20 nodes; (b) the corresponding shortest paths graph and cache. Nodes 4 and 13 are chosen as additional cover nodes along with destination Nodes 14 to 19. Local cache shows the all-pair shortest distance matrix between cover nodes and destination nodes.}
\label{fig:example_cache}
\end{figure}

The greedy approach for Hitting-Set can be implemented with a running-time of $\mathcal{O}(n)$, where $n=\sum_{u=1}^{|V|} |p_u| $. The algorithm maintains a priority queue of the frequencies of the nodes that appear over the uncovered sets. At each iteration, the most frequent node is popped and all the sets that contain that node are removed. For each removed set, each node in the set is examined and the update operation is called for the priority queue to decrement that node's priority. Notice that the update operation has an amortized run-time of $\mathcal{O}(1)$ since the priorities are decremented by at most 1 every time. There are $n$ nodes to be removed and the update operation costs $\mathcal{O}(1)$, hence the run-time of $\mathcal{O}(n)$.

\subsubsection{Caching}
We store $cover(u)$ and $dist(u, cover(u))$ for all of $u \in V$, and the all-pair shortest distances between the cover nodes $C$ and destination nodes $D$. We define this as a partition in the cache. A partition is able to accommodate all distance queries $dist(u, v)$, for $u \in V, v \in D_i$, where $D_i$ is a set of destination nodes. Figure~\ref{fig:example_cache} shows how a cache partition is stored with respect to a small road network. As an example, the query $dist(2, 17)$ would first determine that the cover node for node 2 is 13 and $dist(2, 13)=9$. Then it finds $dist(13, 17)=4$ in the local cache and lastly calculates $dist(2, 17)=dist(2, 13)+dist(13, 17)=9+4=13$. Both $dist(2, 13)$ and $dist(13, 17)$ are $\mathcal{O}(1)$ lookup. All queries can be executed in $\mathcal{O}(1)$ in this fashion.

In order to accommodate distance queries between all pairs of nodes, $k$ partitions are chosen such that ${\bigcup_{i=1}^{k} D_i = V}$, where $D_i$ is the set of destination nodes for partition $i$.  According to our heuristic, the destination nodes of a partition should reside in the same geographical region in order to better leverage the common sub-path property. For our experiments, we used NYC taxi zones as our geographical regions. In addition, we also store the symmetrical version of the cache where each partition accounts for a set of source nodes to all destination nodes. This enables cache-coherent look-ups to speed up the simulation.

\subsubsection{Improvements}
The Manhattan graph has 9888 vertices and 16148 edges while the New York City graph has 140984 vertices and 247268 edges, which is about 14 times larger. Using this caching scheme over the map of Manhattan reduces the size of the cache from 1.5GB to 0.64GB (including both source and destination partitions), which would be insufficient over the New York City graph. Most of the memory is used in storing the shortest distance matrices between the cover nodes and destination nodes, meaning that the resulting sets from the Hitting-Set algorithm were relatively large. Further investigation showed that this was due to a significant number of trivial common sub-paths (sub-paths of size 1).

To further reduce memory usage, we devised a method to extend the sub-paths for the Hitting-Set algorithm. To do so, we extended the definition of cover nodes from a single node to a set of nodes. This allows the search for cover nodes to go further and thus results in a set of longer sub-paths. The additional requirement is that this set of sub-paths should cover every shortest path. To obtain this set of sub-paths, we extend the common sub-path along the paths that it forks into. Each fork is now its own sub-path and together they make
the set of sub-paths. The set of nodes that covers this set of sub-paths will be the cover set. Note that this guarantees an improvement in the results of the Hitting-Set algorithm as all these sub-paths can be "hit" by any node from the single common sub-path. This forking procedure can be performed recursively to achieve even longer sub-paths.
\begin{figure}[t]
\centering
\subfloat[]{
    \includegraphics[width=0.24\textwidth, keepaspectratio]{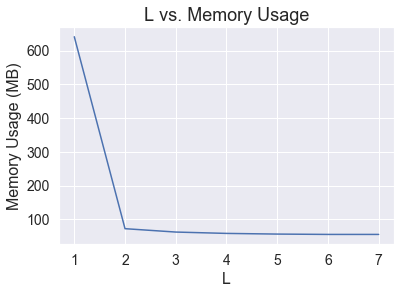}
    \label{fig:t_vs_mem_line_chart}
}
\subfloat[]{
    \includegraphics[width=0.24\textwidth, keepaspectratio]{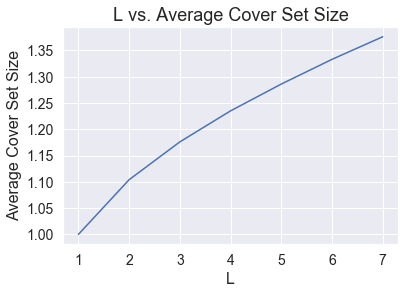}
    \label{fig:t_vs_cover_line_chart}
}
\caption{The cache size (a)
and average cover set size (b) plotted against L (minimum sub-path threshold).}
\label{fig:cache_line_charts}
\end{figure}

\subsubsection{Trade-offs}

This results in longer query times, as distance queries are now required to perform a number of look-ups equivalent to the size of its cover set. The average number of look-ups per query can be estimated prior to the simulation using information about the graph. Notice that when a forking procedure is performed, the lengths of the sub-paths are guaranteed to increase by at least 1. Thus, to achieve sub-paths of at least length $L$, at most $L-1$ forking procedures are performed. For any $L$, the maximum size of a cover set is $b^{L-1}$ where $b$ is the maximum branching factor of the road network. The worst case time complexity for distance queries is then $\mathcal{O}(b^{L-1})$. Road networks are sparse in nature due to the geographical constraints, so $b$ is relatively small across all road networks. 

The actual impact on query time can be estimated by calculating the average cover set size, assuming all source-destination pairs are queried uniformly. With historical query data, we can calculate the exact impact by calculating the weighted average. Figure~\ref{fig:t_vs_mem_line_chart} shows the estimated impact of $L$ on cache size and Figure~\ref{fig:t_vs_cover_line_chart} shows the same on average cover set size. For our experiments, we used $L=2$ due to its minimal increase in overall table look-ups and large reduction in memory usage. For the Manhattan graph, memory usage was reduced to 0.072GB with $L=2$. We were able to reduce the cache size by more than $95\%$ while only incurring a $10\%$ increase in average table look-ups. For the NYC graph, the cache size is 4.8GB, which is small enough to support large scale simulations in parallel.

The preprocessing time has also increased due to the need to compute common sub-paths and hitting sets on top of computing the shortest paths for every node. Computing the cache for the NYC graph would require several hours. To speed up preprocessing, multi-threading is leveraged. The cache partitions can be computed independently from each other, thus allowing the preprocessing phase to be fully parallelized. We were able to compute the cache for the NYC graph in under 1 hour.

\subsubsection{Shortest Path Query}

Unlike distance queries, a path query requires a precise list of visited nodes and their respective distances from the source. For this reason, the predecessor matrix cannot be compressed in any way without losing path information. Computing shortest paths during simulation using generic graph search algorithms is too slow and scales poorly in large graphs. To speed up the search, we introduce a pruning step to achieve $\mathcal{O}(n)$ path queries, where $n$ scales with the length of the path.

Using the computed shortest distances, the search space can be pruned effectively to only include nodes that appear on the shortest path. A node $w$ appears on the shortest path from $u$ to $v$ if and only if ${dist(u,w)+dist(w,v)=dist(u,v)}$. Using this observation we can constrain depth-first search to retrieve shortest paths in $\mathcal{O}(n)$. Since every node visited is guaranteed to appear on a shortest path, no backtracking is required during the search, and therefore the path query scales linearly with the number of nodes in the path. Depth-first search is used instead of other breadth-based search algorithms because shortest paths are not unique and only one is required. Using depth-first search and distance pruning, shortest paths can be computed efficiently during simulation.

\section{Experiental Results}
In this section, we describe two sets of experiments designed to evaluate the performance and the scalability of the STaRS+ model. We first provide some information about the datasets and parameters. Then, in section \ref{accuracy_eval}, we compare the performance of STaRS+ against our ILP model by defining a consistent setting between two frameworks. Finally, in section \ref{stars_scale_eval}, we compare STaRS+ with STaRS in terms of scalability.

\subsection{Datasets and Parameters}
For benchmarking, we used the TLC Trip Record Data\footnote{https://www1.nyc.gov/site/tlc/about/tlc-trip-record-data.page} dated on Thursday, September 5, 2013. The dataset contains 461,739 taxi trip requests that originated in the NYC metro area. Each request is described by a vector of 14 variables including driver ID, pickup/dropoff time, coordinates of pickup/dropoff locations, trip distance, number of passengers, and auxiliary information such as payment type, surcharge, tax, tip, and toll amount. For the experiments, we only considered the requests that originated and ended within the bounded region marked by our road network. In preprocessing, we removed the requests that had inconsistent time stamps or geo-coordinates. This led to a total of 391284 requests in Manhattan and a total of 440541 requests in 5 boroughs of New York City. Even though our frameworks are 
extendable to the settings of multiple passengers in a single request, for simplicity of the experiments, we considered all the requests to have only one passenger.

The set of input parameters, including those imposing the constraints, are summarized in table~\ref{input}. Furthermore, the set of metrics listed in Table \ref{Notations} are defined to evaluate the performance of the models and compare different settings.

\begin{table}[ht]
\centering
\resizebox{\linewidth}{!}
{
\begin{tabular}{|l|l|}
\hline
 Notation & Description\\
\hline
$T_{\text{wait}}$ & maximum waiting time between the request time and pickup time  \\
$T_{\text{total}}$ &  total travel delay including waiting time and in-vehicle delay\\
$Capacity$ & default vehicle capacity \\
$|V| $ & fleet size \\
$ S_v $ & driving speed of vehicles\\
$ S_p $ & walking speed of passengers\\
$D_w$ & maximum walking distance passengers are willing to walk to the meeting point\\
$C_{run}$ & computation time for each iteration\\
$\Delta$ & batching time\\
\hline
\end{tabular}
}
\caption{Notations and descriptions of input parameters of frameworks.}
\label{input}
\end{table} 

\begin{table}[ht]
\centering
\resizebox{\linewidth}{!}
{
\begin{tabular}{|l|l|}
\hline
 Notation & Description\\
\hline
Service Rate & counts of served trips over the total number of request\\
$(\overline{t_{\text{wait}}})$ & average waiting time of served requests\\
$\overline{t_{\text{extra}}}$ & average in-vehicle delay\\
$\overline{t_{\text{total}}}$ & average total travel delay\\
$\overline{VMT}$ & average miles vehicle drove during simulation \\
$\overline{d_{hold}}$ & average waiting time of vehicles (measured as distance)  \\
$\overline{occ}$ & average occupancy (total passengers per vehicle in a given point in time)\\

$\overline{w_{\text{pick}}})$ & average walk from source to pick up point \\
$\overline{w_{\text{drop}}})$ & average walk from drop off point to destionation \\
$\overline{w_{\text{total}}}$ & average total walk \\
$\overline{t_{\text{iteration}}}$ & average computation time per iteration \\
$t_{\text{exe}}$& total execution time of simulation \\
BPH & Boarding per Hour \\
\hline

\end{tabular}
}
\caption{Parameters used to evaluate the performance of different frameworks and models.}
\label{Notations}
\end{table} 
\begin{table*}[ht]
\centering
\resizebox{\linewidth}{!}
{\begin{tabular}{clccccccccccccccc}
 \hline
 & 
 \multicolumn{1}{c}{Model} &
 \multicolumn{1}{c}{Service Rate} &
 \multicolumn{1}{c}{$\overline{t_{wait}}$} &
 \multicolumn{1}{c}{$\overline{t_{total}}$ } &
 \multicolumn{1}{c}{$\overline{t_{extra}}$} &
 
 \multicolumn{1}{c}{$\overline{w_{pick}}$} &
 \multicolumn{1}{c}{$\overline{w_{drop}}$} &
 \multicolumn{1}{c}{$\overline{w_{total}}$} &
 
 \multicolumn{1}{c}{$VMT$} &
 \multicolumn{1}{c}{$VMT+\overline{d_{hold}}$} &
 \multicolumn{1}{c}{$\overline{t_{iteration}}$} &
 \multicolumn{1}{c}{$t_{exe}$} &
 \multicolumn{1}{c}{BPH} &
 \multicolumn{1}{c}{$\overline{occ}$} 
 \\
 \hline
 \parbox[t]{4mm}{\multirow{4}{*}{\rotatebox[origin=c]{90}{RP}}} &  &&&&\\
 
& STaRS+                   & 65.91 
& 214.82 & 413.91 & 199.08 & - & - & - & 260.04 & * & - & 63.7 
& 10.75 &1.69 \\
& USTaRS+                  & 70.44 
& 243.07 & 460.50 & 217.42 & - & - & - & 305.92 & * & 0.04 & 133.0 
& 11.48 &1.93\\
& WSTaRS+                  & 70.46 
& 241.23 & 426.05 & 184.82 & - & - & - & 269.70 & * & 0.08 & 255.5 
& 11.49 &1.84\\
& ILP &84.69   & 166.47 & 384.25 & 217.78 & - & - & - & 335.90 & * & 27.61 & 79516.8 
& 13.81&  2.56 \\
 \hline
  \parbox[t]{2mm}{\multirow{4}{*}{\rotatebox[origin=c]{90}{RPMP}}}\\
& STaRS+                    & 81.79 
& 253.70 & 359.53 & 105.82 & 217.85 & 305.82 & 523.65 & 221.25 & 223.86 & - & 1126.94 
&  13.33  & 1.82\\
& USTaRS+                   & 84.78 
& 285.42 & 434.03 & 148.61 & 224.68 & 338.07 & 562.74 & 273.01 & 274.50 & 0.48 & 1400.89 
& 13.82 &1.91 \\
& WSTaRS+                   & 86.51 
& 279.42 & 375.61 & 96.19  & 220.62 & 309.39 & 530.00 & 224.04 &  227.63 & 0.66 & 1902.4 
& 14.10&1.68\\
& ILP & 92.78 & 165.27 & 267.31 & 102.04 & 117.60 & 340.20 & 457.80 & 319.11 & 320.97 & 136.90 & 394272.0
& 15.13 & 2.38  \\
\hline
\end{tabular}}
\caption{Comparing STaRS+ and ILP-method on the Manhattan graph with NYC Taxi data from Sep 5th, 2013. The first half of the table presents the results of the  RP model while the other half shows the results for RPMP.} 
\label{STARS vs PNAS}

\end{table*}


\subsection{ILP-based model vs. STaRS+ at small scale for accuracy evaluation}\label{accuracy_eval}
To compare the performance of STaRS+ and our ILP-based model in terms of performance and computational tractability, we define a simplified setting in which we keep the components of the two models as consistent as possible. The objective of the simulations is to minimize the total travel time of the fleet while maximizing ridership. We assume pickups and dropoffs are instant (i.e., no dwell time), and vehicles stay idle until a rebalancing destination or a trip is assigned to them. The time window $\Delta = 30 sec$ is used for batching. Correspondingly, fleet rebalancing is done every 30 seconds. The STaRS+ experiments were performed on a workstation with Intel(R) Xeon(R) CPU E5-2630 v4 2.20GHz CPUs, 62 GB RAM, 468 GB, while the ILP-based model experiments were conducted on a workstation with 16 Intel(R) Xeon(R) Gold 6244 3.60GHz CPUs, 187 GB RAM.

The Manhattan road network used in these experiments consists of 4411 intersections and 9625 edges extracted from the open street map. On average, there are 29 meeting points around each intersection in this network. Out of 461,739 trips, 391,284 trips originated and ended in Manhattan are used as the request set. A fleet of 1000 vehicles ($|V| = 1000$) was used in the simulations. Vehicles start at midnight from random initial intersections shared between two frameworks. The capacity of vehicles and their default speed is set to 4 ($Capacity = 4$) and 17.5 mph ($S_v = 17.5 mph$) respectively. The value for $S_v$ is set based on 70\% of the NYC street speed limit. The maximum waiting time for passengers can not exceed 5 minutes ($T_{wait} = 5 minutes$), and the maximum total travel delay allowed is 10 minutes ($T_{total} = 10 minutes$). The meeting points are up to 400 meters ($D_w = 400 m$) away from the source and destination of the request, and the preferred walking speed is 1.4 m/s ($S_p = 1.4m/s$). Furthermore, we consider a 5 second computation time ($C_{run} = 5 sec$) which is added to the waiting time of requests.

In the ILP-based model, we employed a 3 seconds timeout for exploring candidate trips for vehicles in \textit{RTV graph} and 20 seconds timeouts in solving the matching ILP and rebalancing LP. Also, we limited the number of candidate vehicles per request in \textit{RV graph} by only considering the best 30 vehicles based on the time to serve each trip. A constraint flexibility factor of 15\%  was used for waiting time and travel delay constraints in the computation of \textit{RV graph} as explained in section~\ref{heuristics}. The Mosek solver~\cite{mosek} was used as the standard solver for solving assignment ILP and re-balancing LP.


Table \ref{STARS vs PNAS} shows the performance of two frameworks in 
RP and RPMP settings. As expected, higher service rates can be seen in the ILP-based model compared to STaRS+ models. These service rate differences are approximately 14\% and 8\% for the RP and RPMP settings respectively. On the other hand, the higher complexity of the ILP-based models results in higher execution times. This makes the STaRS+ models more efficient in real-time execution for both RP and RPMP settings with their reasonable performance.

Though both models of STaRS+, i.e. USTaRS+ and WSTaRS+, have higher service rates, their average waiting time are over 30 seconds less than the batched versions. It makes STaRS+ computationally more efficient. Since USTaRS+ does not minimize the total travel time of the fleet, it has relatively poor performance compared to other models with the highest values for $t_{wait}$, $t_{extra}$, and $t_{total}$. With respect to the average walking distance to pickup ($p_{walk}$) and the vehicle mile traveled ($VMT$), USTaRS+ is also less efficient despite its similar service rate to WSTaRS+.

As expected, the RPMP model has a higher service rate and a lower $VMT$ in comparison with RP. In terms of performance parameters, the ILP-based models have lower average waiting times ($t_{wait}$), and consequently, lower average pick-up walks ($w_{pick}$) compared to the STaRS+ models. However, the ILP-based models have higher $VMT$ values for a number of reasons. First, the higher service rates in ILP-based models mean that the vehicle fleet has to travel more miles to serve the additional sets of requests. This imposes more challenges and thus contributes more to the higher $VMT$. Secondly, as we mentioned earlier, since ILP-based models have lower $p_{walk}$ values in comparison with STaRS+ models, vehicles may have to drive more to pick up passengers. The vehicle pruning strategy could be one reason why it favors closer vehicles over farther ones and misses potential vehicles that could serve the request with fewer detours. STaRS+, on the other hand, searches for vehicles that can serve the request with the lowest additional cost (time) regardless of their current position. Thus, in many cases, STaRS+ asks passengers to walk for almost $D_w$ to get a ride.
\begin{figure}[t]
\centering
\subfloat[]{
    \includegraphics[width=0.24\textwidth]{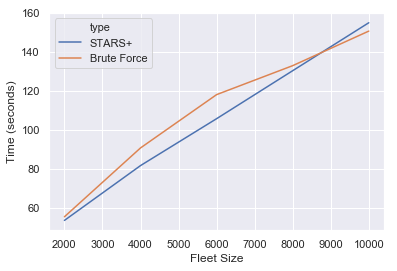}
    \label{fig:cabs_vs_time_line_chart}
}
\subfloat[]{
    \includegraphics[width=0.24\textwidth]{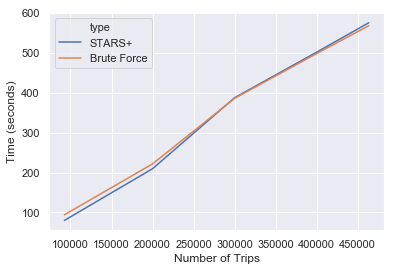}
    \label{fig:trips_vs_time_line_chart}
}
\caption{Comparing the simulation time between STaRS+ caching scheme and brute force distance matrix caching. (a) uses 10000 trips for the experiment and (b) uses a fleet size of 4000. All other parameters were kept constant.}
\label{fig:scalability_line_charts}
\end{figure}
\subsection{STaRS+ Scalability Evaluation}\label{stars_scale_eval}
To demonstrate the scalability and robustness of the STaRS+ caching scheme, we compared it against the previous brute force cache method over the graph of New York City. Brute force caching gives the best theoretical run-time for distance queries, but it is impractical due to its $\mathcal{O}(n^2)$ memory complexity. We demonstrate that our method solves the problem with memory usage while remaining competitive in query time. Simulations were performed on a Linux machine (64 Intel(R) Xeon(R) CPU E5-4640 @ 2.40GHz; 1TB RAM). Caches are fully loaded onto the main memory at the beginning of the simulation. We measure simulation time with respect to varying fleet sizes and trip volumes, while all other parameters are kept constant. For the sake of simplicity, the scalability experiments employ the simpler RP model. Our method consumes 4.8GB while the brute force cache takes 297GB. Figure \ref{fig:scalability_line_charts} displays the results of the experiments. The brute force method is expected to outperform the STaRS+ method by a small factor. However, in practice, the STaRS+ method performed on-par with the brute force method. We speculate the cause to be the increased number of cache misses due to the large cache size. The results empirically support our theoretical analysis of the shortest-distance and shortest-path queries.

\section{Case Study: ride-pooling for all 5 NYC boroughs}\label{casestudy}

\begin{figure}[t]
\centering
\subfloat[]{
    \includegraphics[width=0.24\textwidth, keepaspectratio]{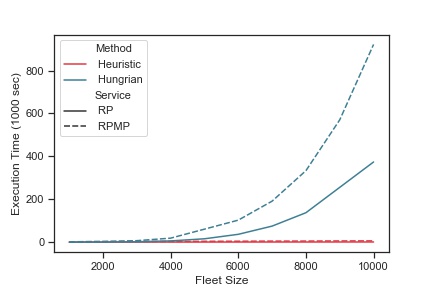}
    \label{fig:heu-vs-hung-exe}
}
\subfloat[]{
    \includegraphics[width=0.24\textwidth, keepaspectratio]{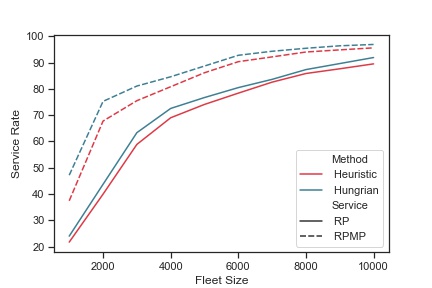}
    \label{fig:heu-vs-hung-service}
}
\caption{The execution time (a) and service rate (b) of RP and RPMP services with STaRS+ and  batch STaRS+ (weighted) plotted against fleet size.}
\label{fig:heu-vs-hung}
\end{figure}

In this section, we evaluate the scalability of STaRS+ by running experiments on five boroughs of New York City. We use the same setting as section~\ref{accuracy_eval} except for the road network. The NYC metro area's road network consists of 140,984 intersections and 247,268 road segments. In this network, there are 67 meeting points around each intersection on average. Tables \ref{table:fiveboroughs-RP} and \ref{table:fiveboroughs-RPMP} show the results of the experiments with RP and RPMP respectively. Figure \ref{fig:heu-vs-hung} on the other hand, compares the performance of STaRS+ and WSTaRS+ models at city scale. WSTaRS+ is almost cubic and as it is plotted in Figure \ref{fig:heu-vs-hung-service}, its slightly higher service rate does not worth its significantly higher execution time.

Furthermore, we evaluated the reliability of the STaRS+ 
by running experiments with various fleet sizes on different occasions during the year in the New York metro area. We used the trip data of a collection of dates that could reflect the pattern of trip requests in extreme situations from least crowded days such as Christmas day or thanksgiving to super crowded days during September or New Year.

\begin{table}[b]
\centering
\resizebox{\columnwidth}{!}
{
\begin{tabular}{lllllllll}
\hline
$|V|$ & 
Service Rate 
& $\overline{t_{wait}}$  
& $\overline{t_{total}}$  &  $\overline{t_{extra}}$  & $VMT$  & 
$t_{exe}$ \\
\hline
1K & 21.74 
& 202.6 & 425.0 & 222.4  & 191.5 & 103.0 \\
2K & 39.9 
& 204.4 & 422.0 & 217.5  & 156.6 & 209.3 &  \\
3K & 58.87 
& 206.2 & 413.4 & 207.1  & 137.8 & 305.8 \\
4K & 69.00 
& 204.5 & 403.5 & 199.0 & 115.8 & 387.4 \\
5K & 74.07 
& 202.6 & 397.8 & 195.2 & 97.4 & 457.1 \\
6K & 78.36 
& 201.7 & 394.7 & 193.0  & 84.5 & 520.3 \\
7K & 82.54 
& 200.6 & 392.1 & 191.5  & 75.4 & 593.8 \\
8K & 85.83 
& 199.7 & 389.6 & 189.9  & 67.9 & 645.1 \\
9K & 87.62 
& 198.9 & 388.0 & 189.1  & 61.0 & 704.1 \\
10K & 89.49 
& 198.0 & 385.6 & 187.6  & 55.5 & 756.4 \\
\hline
\end{tabular}}
\caption{STaRS+ RP on NYC metro area using 1k-10k vehicles. Description of columns is presented in Table \ref{Notations}.} 
\label{table:fiveboroughs-RP}

\end{table} 
\begin{table}[b]
\centering
\resizebox{\columnwidth}{!}
{
\begin{tabular}{llllllllll}
\hline
$|V|$ & 
\multicolumn{1}{p{0.4cm}}{\centering Service\\ Rate}
& $\overline{t_\text{wait}}$ 
& 
$\overline{t_{total}}$  &  $\overline{t_{extra}}$  & $\overline{w_{pick}}$&$\overline{w_{drop}}$& $\overline{w_{total}}$& $VMT$  & 
$t_{exe}$ \\
\hline
1K & 37.45 
& 251 & 395 & 143 & 211.8 & 329.5 & 541.3 & 228.8 & 1115 \\
2K & 67.74 
& 251 & 372 & 120 & 217.5 & 320.9 & 538.5 & 167.6 & 2103 \\
3K & 75.49 
& 249 & 362 & 113 & 216.4 & 319.6 & 536.1 & 119.3 & 2586\\
4K & 80.80 
& 248 & 358 & 110 & 216.7 & 319.1 & 535.8 & 93.4 & 2895 \\
5K & 86.11 
& 247 & 354 & 107 & 216.5 & 318.3 & 534.8 & 77.8 & 3501 \\
6K & 90.32 
& 246 & 351 & 105 & 216.2 & 318.0 & 534.3 & 66.7 & 4003 \\
7K & 92.14 
& 245 & 349 & 103 & 216.2 & 317.8 & 534.0 & 57.5 & 4151 \\
8K & 93.99 
& 244 & 347 & 102 & 215.4 & 317.6 & 533.1 & 50.7 & 4907 \\
9K & 94.81 
& 243 & 345 & 102 & 215.1 & 317.5 & 532.7 & 45.1 & 5094 \\
10K & 95.57 
& 242 & 343 & 101 & 214.2 & 317.2 & 531.5 & 40.6 & 5791 \\
\hline
\end{tabular}}

\caption{STaRS+ RPMP on NYC metro area using 1k-10k vehicles. Description of columns is presented in Table \ref{Notations}.}
\label{table:fiveboroughs-RPMP}

\end{table} 

\begin{figure}[t]
\centering
\subfloat[]{
    \includegraphics[width=0.24\textwidth, keepaspectratio]{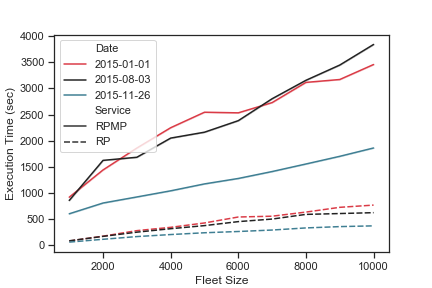}
    \label{fig:3dayexe}
}
\subfloat[]{
    \includegraphics[width=0.24\textwidth, keepaspectratio]{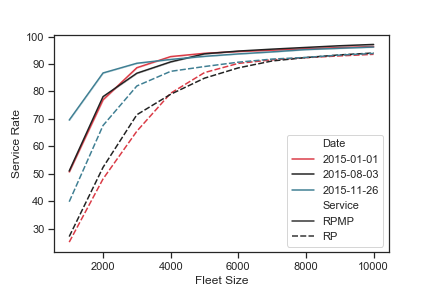}
    \label{fig:3dayservice} 
}
\caption{The Execution time (a) and service rate (b) of RP and RPMP services plotted against fleet size on three different datasets.}
\label{fig:3daycompare}
\end{figure}

The fleet size ($|V|$) and the request set ($R$) are the variants of this set of experiments. TLC Taxi Data of November 26, 2015, with a total of 225,679 trips is used as a representative of relatively less crowded days while September 3, 2015's, and January 1, 2015's records, with 323,918 and 353,265 requests represent above-average and super crowded days in NYC. 
Figure~\ref{fig:3daycompare} displays the execution time and service rate of both RP and RPMP services of the STaRS+ model against the fleet size on these data sets.  

To demonstrate the importance of 
city-scale simulations in decision making, we further present a detailed comparison of the performance of STaRS+ in the Manhattan and NYC metro area. To this end, an extensive set of experiments with varying maximum walking distance ($D_w$), capacity, and maximum waiting time-total delay pair ($T_{wait}, T_{total}$) is designed and ran on both Manhattan and city scales. 

\begin{table}[]
    \centering
    \resizebox{2.6 in}{!}{
    \begin{tabular}{l|cc}
      \hline
    Borough & \#requests & percentage \\
     \hline
    Bronx & 2054 & $00.69\%$\\
    Staten Island & 3 & $0.001007\%$\\
    Brooklyn & 22112 & $7.42\%$ \\
    Queens & 27457 & $9.22\%$ \\
    Manhattan & 246029 & $82.65\%$ \\
     \hline
   \end{tabular}
   }
    \caption{Distribution of requests in boroughs of NYC.}
    \label{tab:distribution}
\end{table}

In most data sets, more than 90\% of requests belong to Manhattan. Therefore, the performance of the models in Manhattan dominates other boroughs in all evaluation metrics. To alleviate this effect, we used combined green and yellow Taxi Data from TLC dated September 7, 2015, in these experiments. The distribution of requests in different boroughs is presented in Table~\ref{tab:distribution}.

\begin{figure}[t]
\centering
\subfloat[]{
    \includegraphics[width=0.24\textwidth, keepaspectratio]{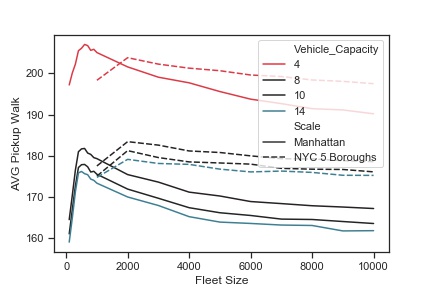}
    \label{fig:case1-a}
}
\subfloat[]{
    \includegraphics[width=0.24\textwidth, keepaspectratio]{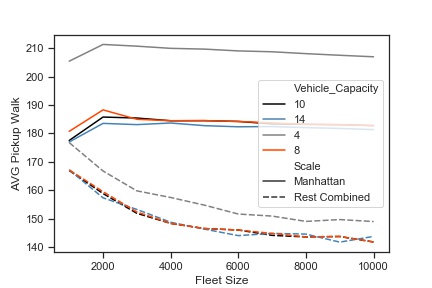}
    \label{fig:case1-b}
}
\caption{(a) The average pickup walk in Manhattan and NYC 5 boroughs against fleet size($|V|$) with varying vehcile \textit{capacity}.
(b) The average pickup walk in Manhattan and other boroughs combined against fleet size($|V|$) with varying vehcile \textit{capacity}.}
\label{fig:case1}
\end{figure}

\begin{figure}[t]
\centering
\subfloat[]{
    \includegraphics[width=0.24\textwidth, keepaspectratio]{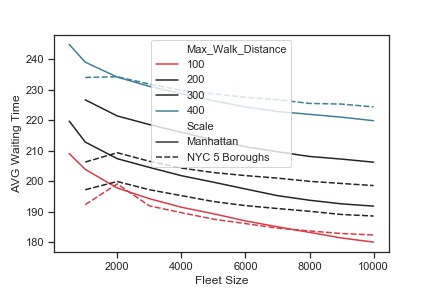}
    \label{fig:case2-a}
}
\subfloat[]{
    \includegraphics[width=0.24\textwidth, keepaspectratio]{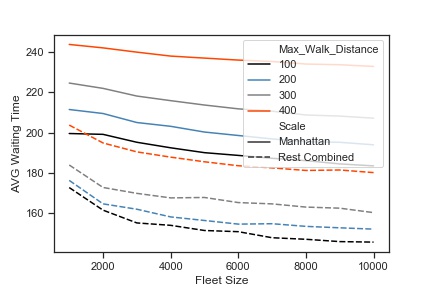}
    \label{fig:case2-b}
}
\caption{(a) The average waiting time in Manhattan and NYC 5 boroughs against fleet size($|V|$) with varying maximum walking distance $D_w$.
(b) The average waiting time in Manhattan and other boroughs combined against fleet size($|V|$) with varying maximum walking distance $D_w$.}
\label{fig:case2}
\end{figure}

%
Furthermore, we split the results of Manhattan from the other four boroughs to compare the patterns. The results of these experiments revealed that though the patterns of Manhattan-scale and city-scale experiments are similar, we cannot come up with a scaling factor to predict the behavior of the NYC metro area based on Manhattan-scale simulations. Considering the waiting time and pickup walk as an example, Figures~\ref{fig:case1}
 and ~\ref{fig:case2} compare these two evaluation metrics for different scales with varying vehicle capacity and walk distance, respectively. ~\ref{fig:case1-a} and ~\ref{fig:case2-a}  compare the performance of Manhattan against NYC metro area and ~\ref{fig:case1-b} and ~\ref{fig:case2-b} decompose the city-scale simulation's results to Manhattan and other boroughs combined. In both cases, the city-scale (other boroughs) line changes more smoothly. Based on ~\ref{fig:case1-b}, walking distance for other boroughs decreases with increasing fleet size while the Manhattan line elevates at first and then decreases with a slower slope in comparison to other boroughs. In Figure~\ref{fig:case2}, a similar difference is evident for the average waiting time.
\section*{Conclusion}
In this paper, we presented STaRS+, a computationally efficient ride-pooling with meeting points framework that deploys a heuristic optimization strategy together with a novel shortest-path caching scheme. We evaluated this framework against our comprehensive ILP-based model. STaRS+ demonstrated comparable performance with significantly higher efficiency.
In order to achieve good performance despite having larger networks, we proposed a caching mechanism which enables constant time shortest path queries while using substantially less memory, but at the cost of additional pre-processing time. Empirically, our caching scheme demonstrated on-par query times with the brute force method. On the other hand, the framework's efficiency is further improved through the use of possible meeting points, which led to a decrease in the vehicles' miles driven. We chose the NYC metro area as a case study to prove our conjecture regarding the necessity of city-scale simulations in urban planning and evaluate the scalability of STaRS+ at once. The results of the experiments revealed that city-scale simulations provide more reliable insight for decision-making by city-planners. Incorporating the pickup/drop off delays and improving the rebalancing strategy to more complicated data-driven models is an interesting future work that could help along with simulations of even more realistic scenarios.

\bibliographystyle{IEEEtran}
\bibliography{paper}
\begin{IEEEbiography}[{\includegraphics[width=1in,height=1.25in,clip,keepaspectratio]
{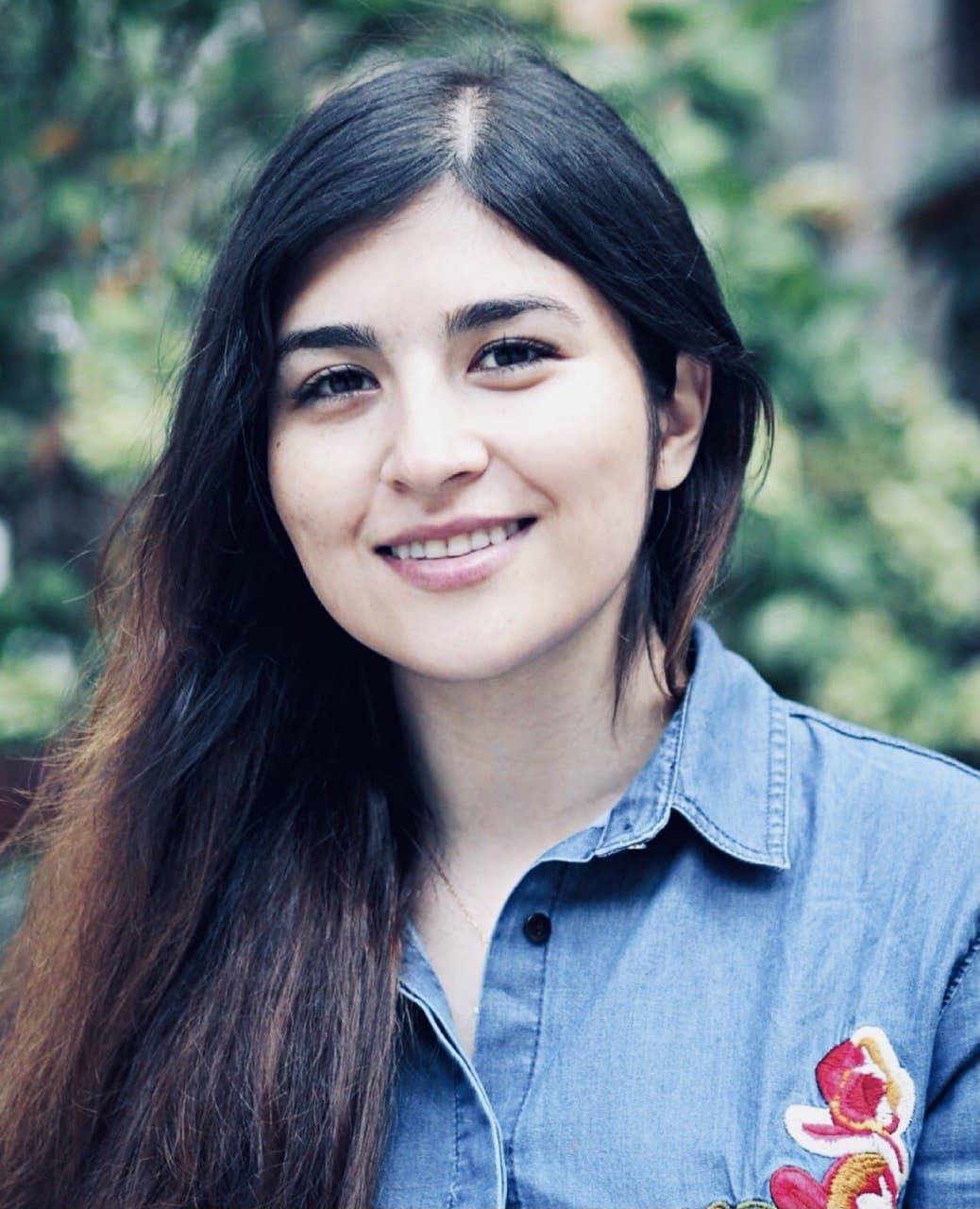}}]{Motahare Mounesan} is a Ph.D. student at the Graduate Center, City University of New York, and a member of the Big Data Interaction Lab at the Center for Urban Science and Progress, New York University. She is also an Adjunct Lecturer at the City College of New York. She received her bachelor's degree in computer science from the University of Tehran, Iran, in 2016. Her research interests include big data management and analytics with a current focus on the partitioning of big multidimensional spatio-temporal data.
\end{IEEEbiography}

\begin{IEEEbiography}[{\includegraphics[width=1in,height=1.25in,clip,keepaspectratio]
{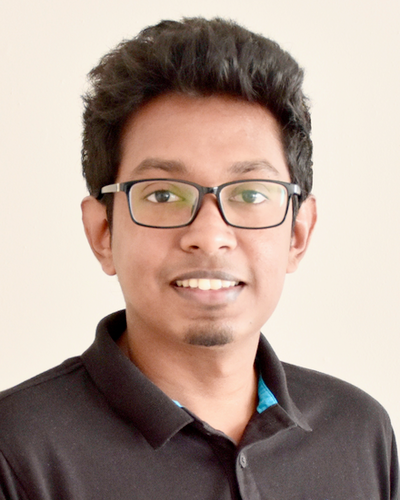}}]{Vindula Jayawardana} is a Ph.D. student at Laboratory for Information \& Decision Systems at Massachusetts Institute of Technology. He received his bachelor’s degree in computer science and engineering from the University of Moratuwa, Sri Lanka. His research interests lie in the intersection of machine learning and optimization with applications into large-scale transportation systems. His recent research focuses on scalable representation learning for mixed autonomy systems in mobility.   
\end{IEEEbiography}

\begin{IEEEbiography}[{\includegraphics[width=1in,height=1.25in,clip,keepaspectratio]
{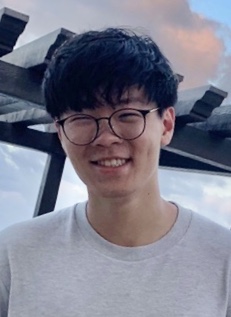}}]{Yaocheng Wu}
is a Ph.D. student at the Graduate Center, City University of New York, and a member of the Big Data Interaction Lab at the Center for Urban Science and Progress, New York University. He received his bachelor's degree in computer science from the City College of New York, in 2019. His research interests include big data visualization, interactive graphics, and vision based systems. His recent research focuses on large point-cloud visualizations.
\end{IEEEbiography}

\begin{IEEEbiography}[{\includegraphics[width=1in,height=1.25in,clip,keepaspectratio]{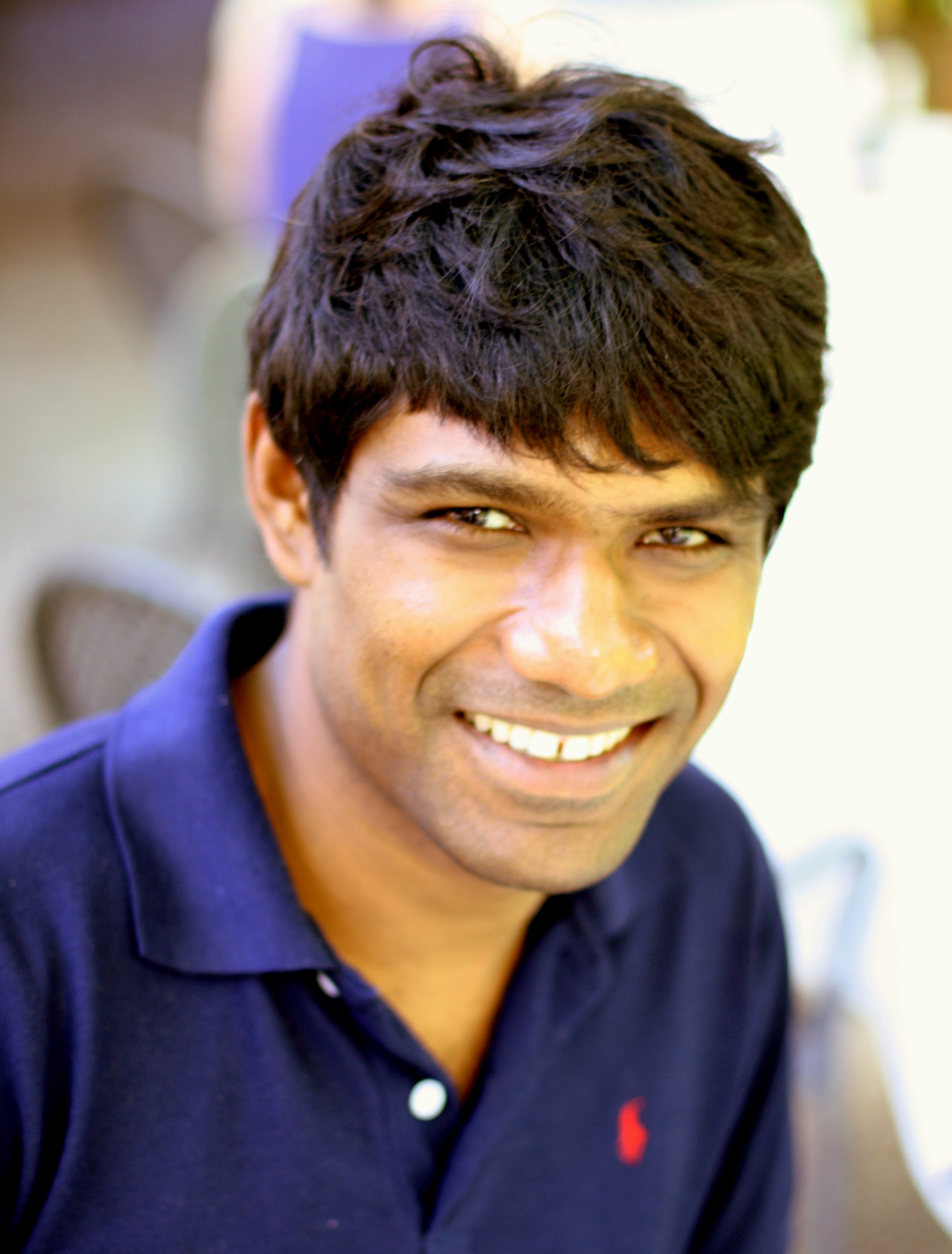}}]{Samitha Samaranayake}
is an Assistant Professor in the School of Civil and Environmental Engineering at Cornell University with graduate field affiliations in Operations Research and Information Engineering, the Center for Applied Mathematics, and Systems Engineering. His research interests are in the modeling, analysis and control of networked urban infrastructure systems with a focus on transportation networks. He received his Bachelors and M.Eng. in Computer Science from MIT, an M.Sc. Management Science and Engineering from Stanford University, and PhD in Systems Engineering from the University of California Berkeley.
\end{IEEEbiography}

\begin{IEEEbiography}[{\includegraphics[width=1in,height=1.25in,clip,keepaspectratio]
{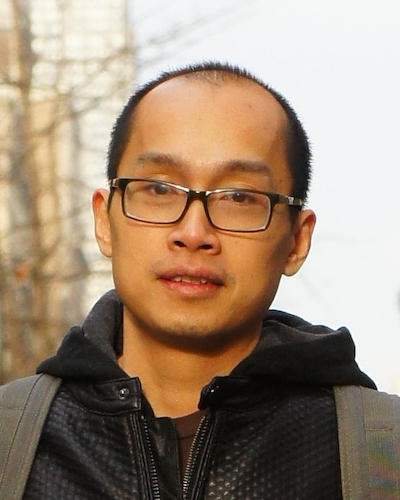}}]{Huy T. Vo}
Huy T. Vo is an Assistant Professor of Computer Science at the City College of New York and a member of the doctoral faculty at the Graduate Center, City University of New York. He is also a faculty member at the Center for Urban Science and Progress, New York University. His current research focuses on high performance systems for interactive visualization and analysis of big data sets, specifically in urban applications. He has co-authored over 50 technical papers and 3 patents, and contributed to several widely-used open-source systems.
\end{IEEEbiography}

 \onecolumn
 \appendix
 \setcounter{page}{1}
 \setcounter{table}{0}
 \begin{table}[ht]
     \centering
     \begin{tabular}{c|cc}
    & Manhattan VS NYC five boroughs & Manhattan VS other boroughs combined\\
    \hline
    \rotatebox[origin=l]{90}{Service Rate} & \includegraphics[width=0.3\textwidth]{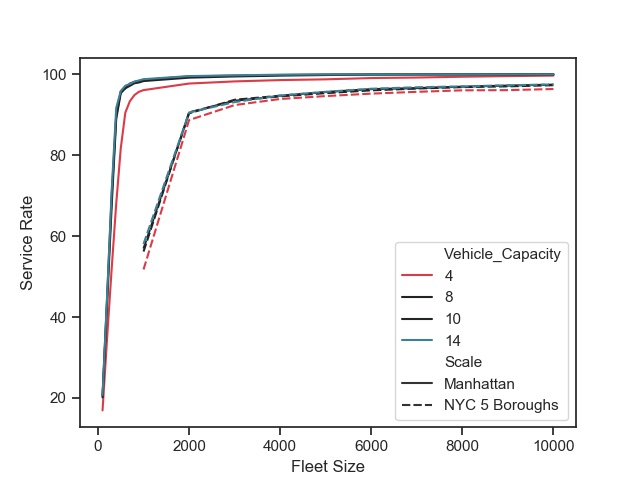}  & \includegraphics[width=0.3\textwidth]{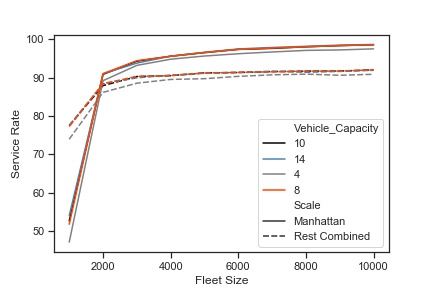}  \\
    \hline
    \rotatebox[origin=l]{90}{AVG Waiting Time} & \includegraphics[width=0.3\textwidth]{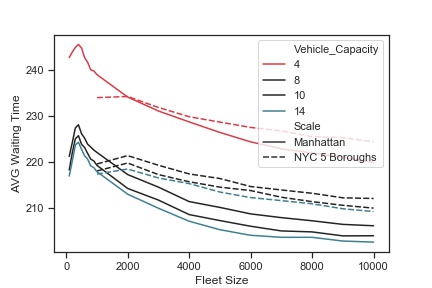}  & \includegraphics[width=0.3\textwidth]{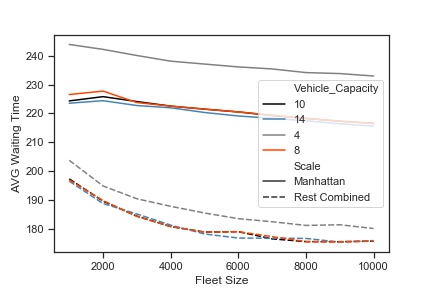}  \\
    \hline
    \rotatebox[origin=l]{90}{AVG In-car Delay} & \includegraphics[width=0.3\textwidth]{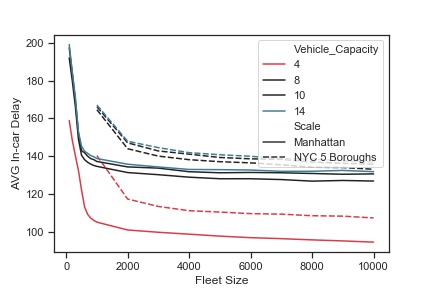}  & \includegraphics[width=0.3\textwidth]{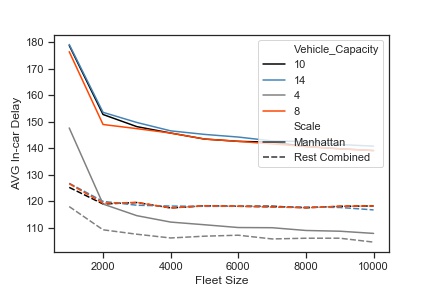}  \\
    \hline
    \hline
    \rotatebox[origin=l]{90}{AVG Pickup Walk} & \includegraphics[width=0.3\textwidth]{figures/Appendix/pickup_fleet_CAPACITY.jpg}  & \includegraphics[width=0.3\textwidth]{figures/Appendix/sep_pickupwalk_fleet_CAPACITY.jpg}  \\
    \hline
    \rotatebox[origin=l]{90}{AVG Dropoff Walk} & \includegraphics[width=0.3\textwidth]{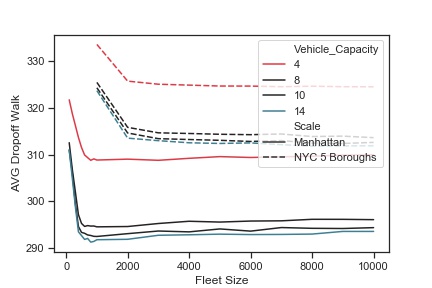}  & \includegraphics[width=0.3\textwidth]{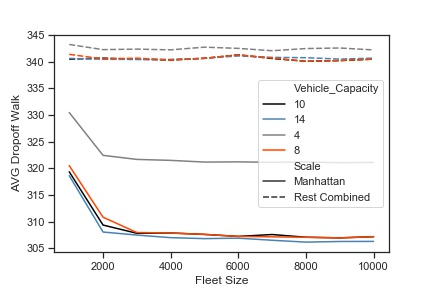}  \\
    \hline
\end{tabular}
     \caption{The first column shows the performance of STaRS+ ride pooling with meeting points (RPMP) in Manhattan and NYC 5 boroughs against fleet size($|V|$) with varying vehcile \textit{capacity}. The second column represent the result of similar experiment for Manhattan and other boroughs combined.}
 \end{table}
 
  \begin{table}[ht]
     \centering
     \begin{tabular}{c|cc}
    & Manhattan VS NYC five boroughs & Manhattan VS other boroughs combined\\
    \hline
    \rotatebox[origin=l]{90}{Service Rate} & \includegraphics[width=0.3\textwidth]{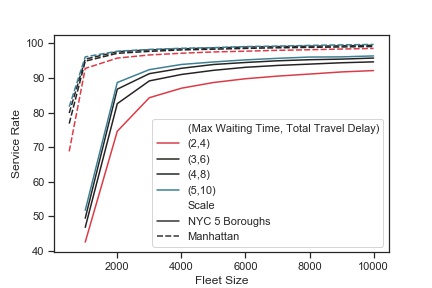}  & \includegraphics[width=0.3\textwidth]{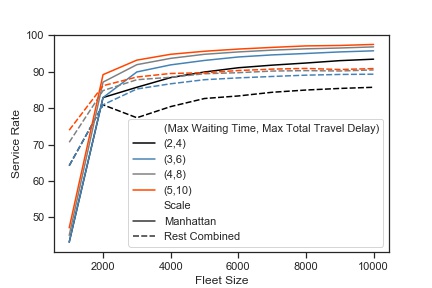}  \\
    \hline
    \rotatebox[origin=l]{90}{AVG Waiting Time} & \includegraphics[width=0.3\textwidth]{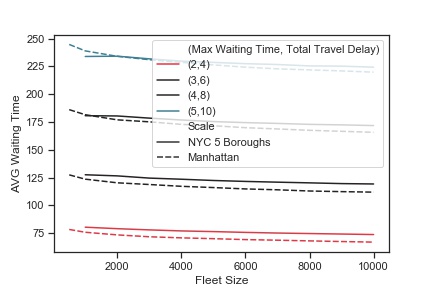}  & \includegraphics[width=0.3\textwidth]{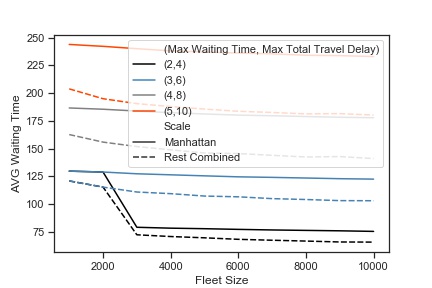}  \\
    \hline
    \rotatebox[origin=l]{90}{AVG In-car Delay} & \includegraphics[width=0.3\textwidth]{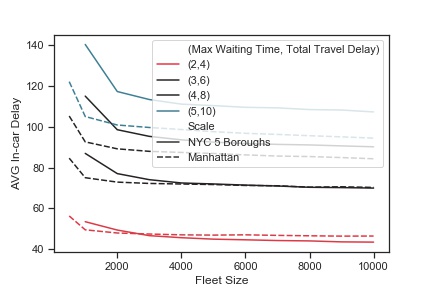}  & \includegraphics[width=0.3\textwidth]{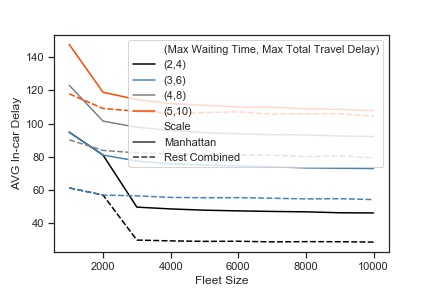}  \\
    \hline
    \hline
    \rotatebox[origin=l]{90}{AVG Pickup Walk} & \includegraphics[width=0.3\textwidth]{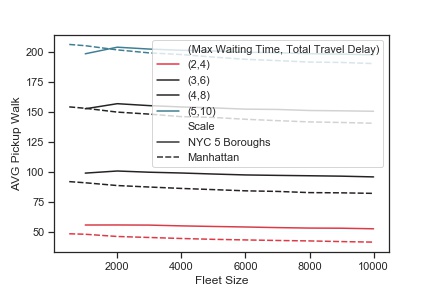}  & \includegraphics[width=0.3\textwidth]{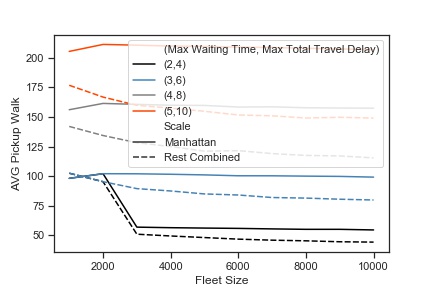}  \\
    \hline
    \rotatebox[origin=l]{90}{AVG Dropoff Walk} & \includegraphics[width=0.3\textwidth]{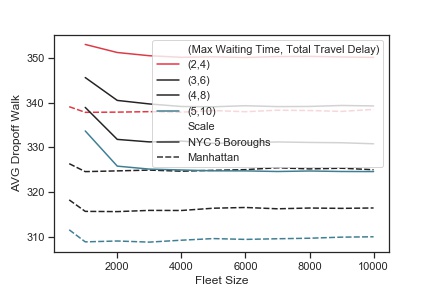}  & \includegraphics[width=0.3\textwidth]{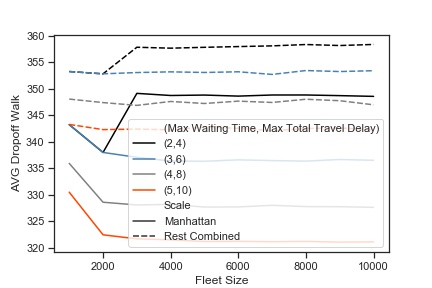}  \\
    \hline
\end{tabular}
     \caption{The first column shows the performance of STaRS+ ride pooling with meeting points (RPMP) in Manhattan and NYC 5 boroughs against fleet size($|V|$) with varying (Waiting Time, Total Delay) pair. The second column represent the result of similar experiment for Manhattan and other boroughs combined.}
 \end{table}
 
   \begin{table}[ht]
     \centering
     \begin{tabular}{c|cc}
    & Manhattan VS NYC five boroughs & Manhattan VS other boroughs combined\\
    \hline
    \rotatebox[origin=l]{90}{Service Rate} & \includegraphics[width=0.3\textwidth]{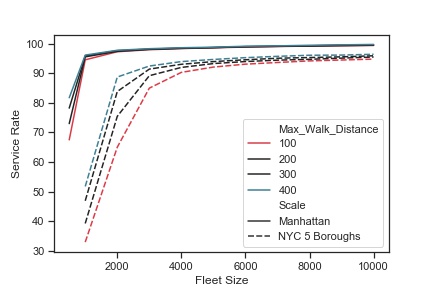}  & \includegraphics[width=0.3\textwidth]{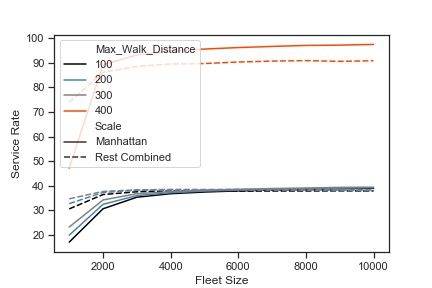}  \\
    \hline
    \rotatebox[origin=l]{90}{AVG Waiting Time} & \includegraphics[width=0.3\textwidth]{figures/Appendix/waittime_fleet_WALK.jpg}  & \includegraphics[width=0.3\textwidth]{figures/Appendix/sep_waittime_fleet_WALK.jpg}  \\
    \hline
    \rotatebox[origin=l]{90}{AVG In-car Delay} & \includegraphics[width=0.3\textwidth]{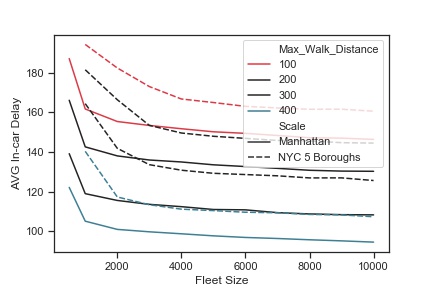}  & \includegraphics[width=0.3\textwidth]{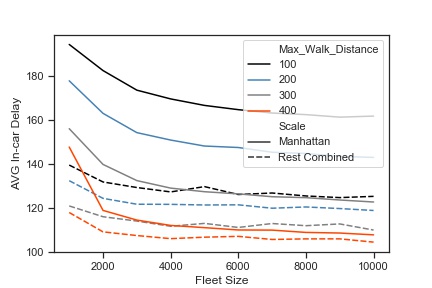}  \\
    \hline
    \hline
    \rotatebox[origin=l]{90}{AVG Pickup Walk} & \includegraphics[width=0.3\textwidth]{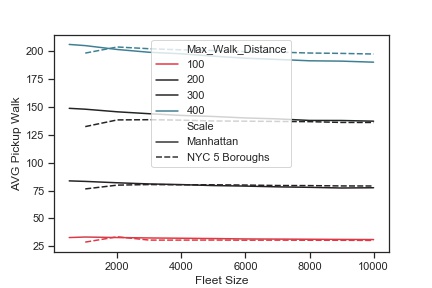}  & \includegraphics[width=0.3\textwidth]{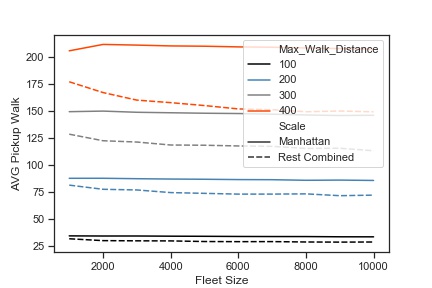}  \\
    \hline
    \rotatebox[origin=l]{90}{AVG Dropoff Walk} & \includegraphics[width=0.3\textwidth]{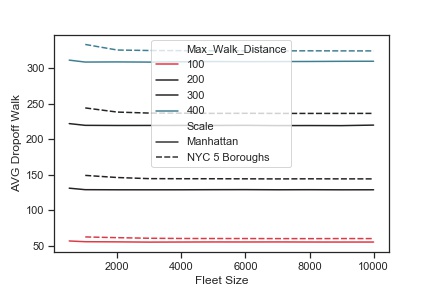}  & \includegraphics[width=0.3\textwidth]{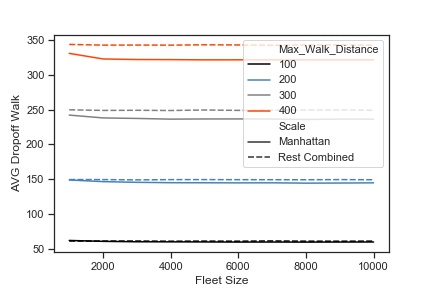}  \\
    \hline
\end{tabular}
     \caption{The first column shows the performance of STaRS+ ride pooling with meeting points (RPMP) in Manhattan and NYC 5 boroughs against fleet size($|V|$) with varying maximum walking distance ($D_w$). The second column represent the result of similar experiment for Manhattan and other boroughs combined.}
 \end{table}

\end{document}